\documentclass[prd,preprint,showpacs,preprintnumbers,amsmath,amssymb]{revtex4}
\usepackage{graphicx}
\usepackage{dcolumn}
\usepackage{bm}

\usepackage{color}

\usepackage{feynmf}
\usepackage{slashed}

\usepackage[utf8]{inputenc}

\begin{document}

\title{Thermally modified sterile neutrino portal dark matter and gravitational waves from phase transition: The Freeze-in case }

\author{Ligong Bian,}
\thanks{lgbycl@cqu.edu.cn}
\affiliation{ Department of Physics, Chongqing University, Chongqing 401331, China}

\affiliation{ Department of Physics, Chung-Ang University, Seoul 06974, Korea}

\author{Yi-Lei Tang}
\thanks{tangyilei@kias.re.kr}
\affiliation{Quantum Universe Center, Korea Institute for Advanced Study, Seoul 02455, Republic of Korea}

\date{\today}

\begin{abstract}

We consider the thermal effects into the evaluation of the dark matter production process.
With the assistance of the right handed neutrinos, the freeze-in massive particle dark matter production history can be modified by the two-step phase transitions. The kinematic of decay/inverse decay or annihilation processes can be affected by the finite temperature effects as the Universe cools down. The history of the symmetry respected by the model can be revealed by the DM relic abundance evolution processes. The strong first order electroweak phase transition generated gravitational waves can be probed. The number of extra scalars for the Hierarchy problem can be probed through the Higgs off-shell searches at the LHC.  

\end{abstract}


\maketitle
\section{Introduction}

The baryon asymmetry of the Universe (BAU) and the dark matter (DM) are two fundamental physics problems that urge the 
particle physicists and cosmology physicists to propose variants of intelligent ideas and models. The electroweak baryogenesis mechanism has been studied extensively to solve the BAU problem due to the detectable signals of 
the strongly first order electroweak phase transition (SFOEWPT) at high energy colliders~\cite{Arkani-Hamed:2015vfh}. 
A SFOEWPT can generate a detectable gravitational wave signal with a typical peak frequency $\mathcal{O}(10^{-3}-10^{-1}) $Hz.
Extensive studies of the gravitational waves generated by the SFOEWPT in various new physics models are inspired especially after the first discovery of the merging black holes with the gravitational waves signals detected by LIGO~\cite{Abbott:2016blz}.  

The existence of the DM is supported by substantial astrophysical and cosmological
observations.
With the accumulation of the direct and indirect detection experimental data of the DM, the Weakly Interacting Massive Particle (WIMP) paradigm confronts an increasing pressure through the interactions between the dark and the visible sectors which are detectable at colliders. The WIMP DM produced at super low temperature around $T_{fo}\sim m_{DM}/26$ where the DM particles decouple from the thermal plasma. An alternative approach is the feebly interacting massive particle (FIMP) production mechanism~\cite{Hall:2009bx}, where the DM particles never reach thermal equilibrium with the SM bath in the whole cosmology history due to the small interaction rate with the Standard Model particles.
The DM particle freeze in occurs at a relatively higher temperature around  
$T_{fi}\sim m_{DM}/x_{fi}$ with $x_{fi}\sim \mathcal{O}(1-5)$ in the FIMP paradigm. 
Considering the SFOEWPT occurs around $T_{n}\sim \mathcal {O}(10\sim 10^3)$ GeV which is indeed around the $T_{fi}$ when the $m_{DM}\sim \mathcal{O}(10\sim 10^3)$ GeV, we can expect that the production of the FIMP DM would be significantly modified by the SFOEWPT. This is due to the
kinematical threshold that can be altered by the thermal correction to the particles that take part in the decay/inverse decay or annihilation processes contributing to the DM abundance production. 

Recently, the Ref.~\cite{Baker:2016xzo,Baker:2017zwx} studied the FIMP DM scenario affected by the thermal masses and the phase transition 
effects. 
 In this work, we present a novel multi-step FIMP production mechanism of the DM abundance after taking into account the SFOEWPT after the reheating of the Universe. In the model to be computed, a pseudo-Dirac sterile neutrino is introduced. In the literature, such kind of sterile neutrinos are utilized to give rise to the neutrino masses through the linear or inverse seesaw mechanisms, with the linear seesaw 
can be differentiated from the inverse seesaw with the feasibility of leptogenesis~\cite{Falkowski:2011xh} and lepton number violation search at colliders. Here, 
we note that the behaviors of the dark matter are not significantly affected by the types of the sterile neutrinos (Dirac or Majorana). See Ref.~\cite{Dib:2016wge,Dib:2017iva,Dib:2017vux} for recent works on collider searches. 
In the dark sector, we introduce a hidden singlet scalar and a hidden fermion. Both these particles can serve as the DM particle depending on the mass spectrum\footnote{For the FIMP DM without thermal effects in the similar model we refer to Ref.~\cite{Becker:2018rve,Chianese:2018dsz,Falkowski:2017uya}.}. 
For completeness, we study the complete set of Boltzmann equations including the thermal effects. The non-thermal production of the FIMP fermionic DM is an excellent benchmark where the gravitational wave signals of SFOEWPT can be reached by the projected GW detectors. The mixing between the active neutrino and the sterile neutrinos may be beyond the colliders search sensitivity. While, the SFOEWPT signals can be reached at LHC, see Ref.~\cite{Goncalves:2017iub}.

\section{The Model}

In this paper, we utilize the model similar to Ref.~\cite{Falkowski:2011xh,Escudero:2016ksa,Falkowski:2017uya,Tang:2016sib}. It contains a $Z_2$-odd majorana fermion $\chi$ and a real-scalar boson $\phi$. The SM-fields are all even under the $Z_2$. We also introduce a $Z_2$-even sterile neutrino. This sterile neutrino can be majorana or pseudo-Dirac. The sterile neutrino-Higgs-lepton boublet couplings in the pseudo-Dirac case can be in a much wider range than the couplings in the majorana case, and the freeze-in process is not much disturbed by details of the sterile-neutrino sector. In this paper, we focus on the pseudo-Dirac case. The general Lagrangian is given by
\begin{eqnarray}
\mathcal{L} &=& \frac{1}{2} \overline{\chi} ( i \gamma^{\mu} \partial_{\mu} - m_{\chi} ) \chi +  \overline{N_D} ( i \gamma^{\mu} \partial_{\mu} - m_{N_D} ) N_D + \frac{1}{2} (\partial^{\mu} \phi \partial_{\mu} \phi - m_{\phi}^2 \phi^2) \nonumber \\
&+& (\mu_1 \overline{N_D^C} P_L N_D + \mu_2 \overline{N_D^C} P_R N_D + \text{h.c.}) + \frac{\lambda_{\phi}}{4} \phi^4 + \lambda_{h\phi} \phi^2 H^{\dagger} H \nonumber \\
&+& (y_{\chi D} \overline{\chi} N_D \phi + i y_{\chi D 5} \overline{\chi} \gamma^5 N_D \phi + y_{N i} \overline{N} P_L l_i \cdot H + y_{N C i} \overline{N^C} P_L l_i \cdot H \nonumber \\
&+& \text{h.c.} ) + \mathcal{L}_{\text{SM}}, \label{PDLag}
\end{eqnarray}
where $\chi^C = \chi$ is the four Dirac four-spinor, 
$l_i$ with $i=1,2,3$ or $i=e, \mu, \tau$ are the SM left-handed lepton doublets.$N_D = \left[ \begin{array}{c} N_1 \\ i \sigma^2 N_2^{*} \end{array} \right]$ is a Dirac four-spinor, and the $N_1$ and $N_2$ are the Weyl-components. $m_D$ is the Dirac mass terms, and $\mu_{1, 2}$ are the Majorana mass terms, $m_{\chi, \phi, N}$ are the $\chi$, $\phi$, $N$ mass terms, and $y_{\chi, \chi 5, N i}$, $\lambda_{\phi, \phi H}$ are the coupling constants. We rotate to a basis that the $y_{\chi, \chi 5, N i}$, $\lambda_{\phi, \phi H}$ and $m_{\chi, \phi, N}$ are real numbers. For simplicity, we omit the $y_{\chi 5}$ which breaks the CP symmetry. In this paper, we adopt the convention in the left-handed lepton and Higgs doublets
\begin{eqnarray}
l_i = \left[ \begin{array}{c} \nu_i \\ e_{L_i}^{-} \end{array} \right],~~~H = \left[ \begin{array}{c} G^+ \\ \frac{v+h + i G^0}{\sqrt{2}} \end{array} \right],
\end{eqnarray}
where $G^{+}$, $G^0$ are the goldstone bosons, h is the SM Higgs boson, and $v = 246 \text{ GeV}$. Note that $A \cdot B = A_i (i \sigma^2_{i j}) B_j$, where $\sigma^2$ is the second Pauli-matrix.

Although in this paper, we only introduce one pseudo-Dirac sterile-neutrino in our calculations, and this seems to be different from some common see-saw models, we can always rotate the mass basis in the $m_{N, D} \propto I$ case so that only one sterile neutrino interact with the dark matter, see Appendix.~\ref{sec:nm}. Therefore, our calculations are still available.
In general, the  ``pseudo-Dirac'' particle with the nonzero $\mu_{1,2}$ and $y_{N C i}$ terms actually split into two nearly-degenerate majorana components. However, these terms are usually rather small in both the linear- and inverse-seesaw models, and their effects in the early universe are usually negligible. Therefore, for convenience, we just set all of them to be zero during the calculation processes.

\section{Electroweak phase transition dynamics}
\label{sec:mvevT}

In this work, we study the effects of a two-stage phase transition on the DM production. The first-stage of the phase transition is a second-order phase transition from the symmetric phase (both $Z_2$ symmetry and the EW symmetry are all preserved) to the phase where the hidden scalar $\phi$ got VEV (where the $Z_2$ symmetry breaks). The second-stage of the phase transition is a first-order phase transition from the $Z_2$ symmetry broken and EW symmetry preserved phase to the $Z_2$ symmetry restored and EW symmetry broken phase. The phase transition can be characterized by the thermal evolution of the thermal potential.
The tree level scalar potential at zero temperature is given by,
\begin{eqnarray}
V_0(h,\phi)=-\frac{\mu^2}{2}h^2+\frac{\lambda_h}{4}h^4+\frac{\mu_\phi^2}{2} \phi^2+\frac{\lambda_\phi}{4}\phi^4+\frac{\lambda_{h\phi}}{2} h^2  \phi^2\;.
\end{eqnarray}
The desired vacuum structure for the two-step phase transition can be obtained easily with the local vacuum localized
$(h,\phi)=(0,v_\phi)$ and the global Electroweak vacuum localized at $(h,\phi)=(v,0)$,
\begin{eqnarray}
\frac{d \,V_0(h,\phi)}{d\, h}\big |_{h=0,\phi=v_\phi}=0\; , ~\frac{d \,V_0(h,\phi)}{d\,\phi}\big |_{h=v,\phi=0}=0\;.
\end{eqnarray}
Which give raise to $\mu^2=\lambda v^2$ and $v_\phi=\sqrt{-\mu_\phi^2/\lambda_\phi}$.
Then, the mass of the hidden scalar $\phi$ is given by: $m_\phi^2=\mu_\phi^2+\lambda_{h\phi}v^2$. 
The existence of a nonzero $v_\phi$ requires
\begin{eqnarray}\label{eq:phivac}
\mu_\phi^2\equiv m_\phi^2-\lambda_{h\phi}v^2<0\;.
\end{eqnarray}
 Firstly, the local vacuum at $(h,\phi)=(0,v_\phi)$ should be higher than the global one at $(h,\phi)=(v,0)$, $V_0(0,v_\phi)-V_0(v,0)>0$, which constrains the parameters as,
\begin{eqnarray}\label{eq:vac}
\lambda_\phi>\frac{2(m_\phi^2-\lambda_{h\phi}v^2)^2}{m_h^2 v^2}\;.
\end{eqnarray}
The electroweak phase transition can be studied in a gauge invariant approach~\cite{Patel:2011th} after taking into account the thermal corrections, the thermal potential used to estimate the vacuum structures at finite temperature is given by,
\begin{eqnarray}
V_T(h,\phi)=-\frac{\mu^2-c_h T^2}{2}h^2+\frac{\lambda_h}{4}h^4+\frac{\mu_\phi^2+c_\phi T^2}{2} \phi^2+\frac{\lambda_\phi}{4}\phi^4+\frac{\lambda_{h\phi}}{2} h^2  \phi^2\;,
\end{eqnarray}
where
\begin{eqnarray}
c_h = \frac{1}{16} (g_1^2 + 3 g_2^2 )+ \frac{1}{4} y_t^2 +  \frac{\lambda}{2}  + \frac{\lambda_{\phi h}}{12}\;, ~
c_{\phi} = \frac{1}{4} \lambda_{\phi} + \frac{1}{3} \lambda_{\phi h}\;. \label{CCPhi}
\end{eqnarray}
with the contribution of fermions $\chi$ negligible since we focus on the freeze-in DM scenarios.

\begin{figure}[!ht]
\begin{center}
\includegraphics[width=0.6\textwidth]{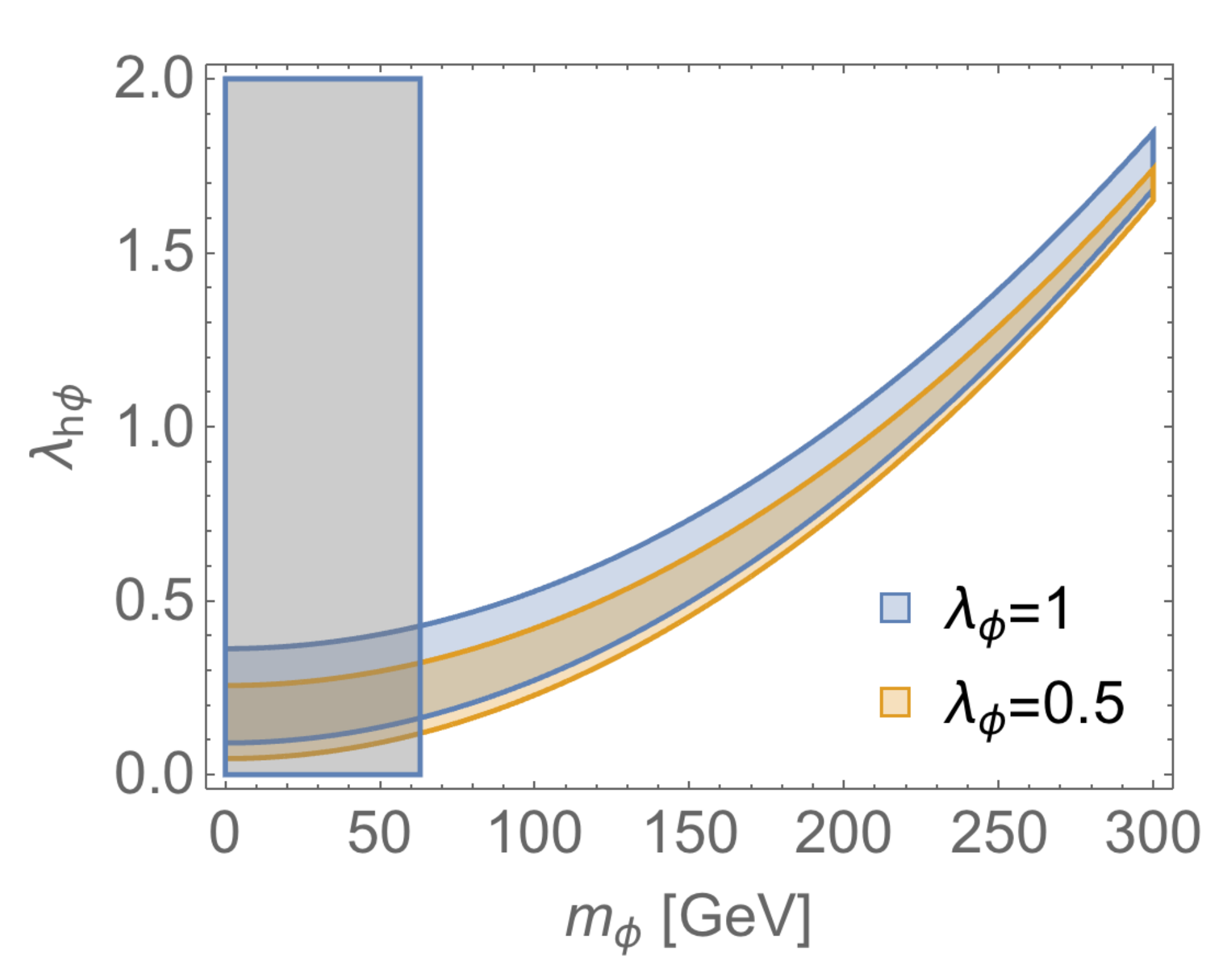}
\caption{Two-step EWPT select conditions allowed parameter spaces. The gray region is excluded by the Higgs invisible decay bounds.}
\label{fig:parewpt}
\end{center}
\end{figure}

To ensure the 
second-stage strong first order phase transition to occur,
 one needs the vacuum in the direction of $\phi$ (at temperature $T_\phi$) to appear earlier than the one in the direction of $h$(at temperature $T_h$) during the phase transition, i.e., $T_\phi>T_h$,
 \begin{eqnarray}\label{eq:Tana}
 \mu^2/c_h<-\mu_\phi^2/c_\phi\;.
 \end{eqnarray}
 Then, the Eq.~\ref{eq:phivac}, Eq.~\ref{eq:vac}, and Eq.~\ref{eq:Tana} together bounds the parameter spaces of $\lambda_\phi,\lambda_{\phi h}$ and $m_\phi$, as shown in Fig.~\ref{fig:parewpt}.

To evaluate the DM at different temperatures, the field dependent mass at zero and finite temperatures are crucial, the zero temperature field dependent mass are,
\begin{eqnarray}
&&m_{hh}=-\mu^2+3\lambda h^2+\lambda_{h\phi}\phi^2\;,\\
&&m_{h\phi}=2\lambda_{h\phi}h\phi\;,\\
&&m_{\phi\phi}=\mu_\phi^2+\lambda_{h\phi}h^2+3\lambda_\phi \phi^2\;,\\
&&m_{G^0}^2=-\mu^2+\lambda_{h\phi}\phi^2+\lambda h^2\;,\\
&&m_{G^\pm}^2=-\mu^2+\lambda_{h\phi}\phi^2+\lambda h^2\;.
\end{eqnarray}
In the Electroweak vacuum, the thermal corrected mass are given by,
\begin{eqnarray}
&&m_{G^0}^2(T)=m_{G^0}^2+c_h T^2\;,\\
&&m_{G^\pm}^2(T)=m_{G^\pm}^2+c_h T^2\;.
\end{eqnarray}
for Goldstones, and
\begin{eqnarray}
m_h^2(T)&=&\frac{1}{2}\bigg(m_{hh}+c_h\,T^2+m_{\phi\phi}+c_\phi \,T^2\mp\sqrt{\big(m_{hh}+c_h\,T^2-(m_{\phi\phi}+c_\phi \,T^2)\big)^2+4 m_{h\phi}^2}\bigg)\;,\nonumber\\
m_\phi^2(T)&=&\frac{1}{2}\bigg(m_{hh}+c_h\,T^2+m_{\phi\phi}+c_\phi \,T^2\pm\sqrt{\big(m_{hh}+c_h\,T^2-(m_{\phi\phi}+c_\phi \,T^2)\big)^2+4 m_{h\phi}^2}\bigg)\;,\nonumber
\end{eqnarray}
for $m_h^2(T)<m_{\phi}^2(T)$ and $m_h^2(T)>m_{\phi}^2(T)$.

The VEVs of the Higgs $h$ and scalar $\phi$ in the $h$-vacuum and the $\phi$-vacuum as a function of temperature are given by,
\begin{eqnarray}
v_h(T)&=&\pm\sqrt{(\mu^2-c_h T^2)/\lambda}\;,\\
v_\phi(T)&=&\pm\sqrt{(\mu_\phi^2-c_s T^2)/\lambda_\phi}\;.
\end{eqnarray} 
Which leads to the variation of particle masses that enter into the DM production process, see the Sec.~\ref{sec:nonDM}.
As the temperature cools down, the scalar masses are essential for the evaluation of the evolution of the DM number density.
Prior to the second stage first order phase transition, all particles are set in the $\phi$-vacuum(wherein the $Z_2$ symmetry is broken and the SM EW symmetry is preserved), and there is no Electroweak symmetry breaking, the scalar field dependent thermal corrected mass are, 
\begin{eqnarray}
\langle m_h^2(T)\rangle _{\phi}&=&m_h^2(T)|_{h=0,\phi=v_\phi(T)}\;,\\
\langle  m_\phi^2(T)\rangle_{\phi}&=&m_\phi^2(T)|_{h=0,\phi=v_\phi(T)}\;,\\
 \langle m_{G^{\pm,0}}^2(T)\rangle_{\phi}&=&m_{G^{0,\pm}}^2(T)|_{h=0,\phi=v_\phi(T)}\;.
\end{eqnarray} 
Assume an instantaneous transition from the electroweak symmetric phase ($\phi$-vacuum) into the electroweak symmetry breaking phase ($h$-vacuum, wherein the $Z_2$ symmetry of $\phi$ is restored and the SM EW symmetry is broken), i.e., a instantaneous supercooling process, we
have the  the scalar field dependent thermal corrected mass being,
\begin{eqnarray}
\langle m_h^2(T)\rangle _{\phi}&=&m_h^2(T)|_{h=v_h(T),\phi=0}\;,\\
\langle  m_\phi^2(T)\rangle_{\phi}&=&m_\phi^2(T)|_{h=v_h(T),\phi=0}\;.\\
 \langle m_{G^{\pm,0}}^2(T)\rangle_{\phi}&=&m_{G^{0,\pm}}^2(T)|_{h=v_h(T),\phi=0}\;.
\end{eqnarray} 
To evaluate the assumption, we calculate the bounce solution to find the  
nucleation temperature of the bubble $T_n$ where the phase transition from the $\phi$-vacuum to $h$-vacuum occurs. 
This makes the 
assumption established for the DM number density evaluation, the final DM number density in the 
$\phi$-vacuum will be taken as the initial number density in the $h$-vacuum.  Therefore, we directly cast the DM number density in the $\phi$-vacuum to the $h$-vacuum around the second stage first order phase transition.

When the Universe cools down to the nucleation temperature $T_n$, which is pretty close to the critical temperature, the phase transition from $\phi$-vacuum to $h$-vacuum proceeds. At this temperature, the bubble nucleation rate per unit volume per within the Hubble horizon $H^{-1}$ reaches unity,
\begin{eqnarray}
\Gamma/V\approx T^4 \exp{-S_3(T)/T}\sim 1\;.
\end{eqnarray}
This condition can be converted to $S_3(T_n)/T_n=4\ln(T_n/H_n)\approx 140-150$~\cite{Apreda:2001us}.
Where $S_3(T_n)$ is the minimized three dimensional Euclidean action evaluating along the bounce configurations. With the
\begin{eqnarray}
S_3(T)=4\pi\int r^2d r\bigg[\frac{1}{2}\big(\frac{d h}{dr}\big)^2+\frac{1}{2}\big(\frac{d\phi}{dr}\big)^2+V(h,\phi,T)\bigg]\;,
\end{eqnarray}
 one can obtain the bounce configuration of the two-fields $h$ and $\phi$ that connects the EW broken vacuum ($h$-vacuum, the true vacuum) and the $Z_2$ broken vacuum ($\phi$-vacuum, the false vacuum) through solving the equation of motion for $h$ and $\phi$,
\begin{eqnarray}
\frac{d^2\phi_b}{dr^2}+\frac{2}{r}\frac{d\phi_b}{dr}-\frac{\partial V(\phi_b)}{\partial \phi_b}=0\;,
\end{eqnarray}
with the boundary conditions of 
\begin{eqnarray}
\lim_{r\rightarrow \infty}\phi_b =0\;, \frac{d\phi_b}{d r}|_{r=0}\;,
\end{eqnarray}
where $\phi_b$ represents the $h_n$ and $\phi_n$. In particular, we adopt the method developed in Ref.~\cite{Cline:1999wi,Profumo:2010kp,Wainwright:2011kj}.


\begin{figure}[!htp]
\begin{center}
\includegraphics[width=0.6\textwidth]{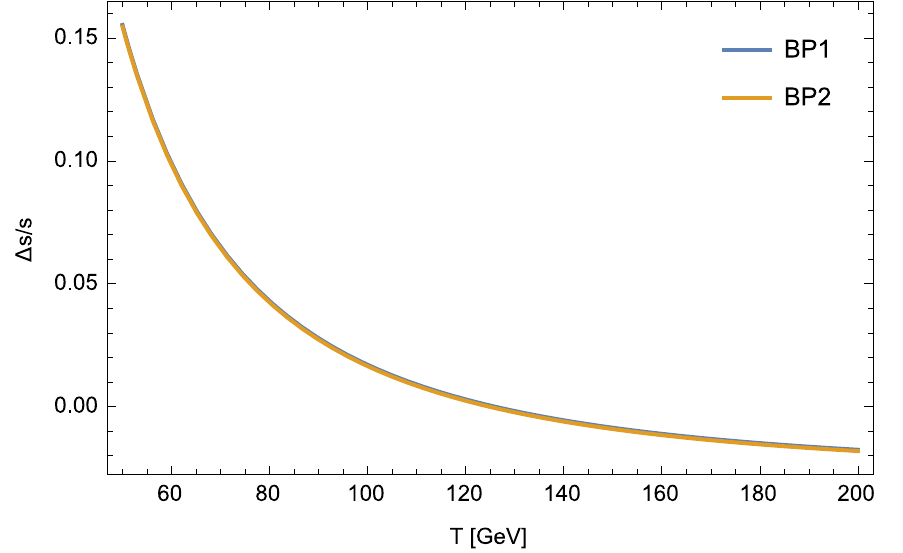}
\caption{The entropy deviation for the benchmarks of Fig.~\ref{fig:FIMPchi} (BP1) and Fig.~\ref{fig:FIMPchiBP3} (BP2).}
\label{fig:EWPTentropy}
\end{center}
\end{figure}
Before the study of the DM in detail we firstly study the entropy induced by the EWPT.
We recall the knowledge of entropy deviation induced by the EWPT following Ref.~\cite{Wainwright:2009mq}.
Assuming that reheating
happens quickly relative to the expansion rate, the energy density $\rho$ of the universe does 
not change during reheating. And there are only small amounts of reheating by the release latent heat of the transition. We do not expect the phase coexistence stage for SFOEWPT.

The injection of the entropy from supercooling process of EWPT for the two step pattern is evaluated by,  
\begin{equation}
\Delta s=-(\frac{dV}{dT}|_\phi-\frac{dV}{dT}|_H)\;,
\end{equation}
with the finite temperature potential $V$ included only the thermal mass corrections evaluated at the nucleation temperatures, which has been normalized by the SM entropy in the radiation dominate universe, 
\begin{equation}
s=\frac{2\pi^2}{45}g_{\star s} T_{EW}^3\;.
\end{equation}
By taking $g_{\star s}=100$, it was estimated that $\Delta s/s_{EW}$ is about percent level around the phase transition temperature $T_n$ for both benchmarks of Fig.~\ref{fig:GW}.
With the entropy difference between the high temperature and low temperature entropy: $\Delta s= s_+-s_-$, one have 
\begin{eqnarray}
\left(\frac{a_f}{a_i}\right)^3 =\frac{1}{1-\Delta s/s_+}=\frac{s_+}{s_-}\;.
\end{eqnarray}
The Fig~\ref{fig:GW} indicates one obtain a negligible dilution factor as in the SM case, see Ref.~\cite{Quiros:1999jp}, and the two-step phase transition pattern being studied in Ref.\cite{Patel:2012pi}.

\section{Thermal effects modified dark matter production}
\label{sec:nonDM}

The calculations of the relic abundance of the thermally modified dark matter are based on the Boltzmann equations (We derive the equations according to Ref.~\cite{LeptogenesisInt, Leptogenesis_NonThermal}). For completeness, we consider both the case of $m_{\phi} > m_{\chi} + m_{N_{(D)}}$ and $m_{\phi} < m_{\chi} + m_{N_{(D)}}$. 
Keeping in mind that, with the temperature cooling down, the phase goes through a first-stage second-order phase transition and then a second-stage first-order phase transition with the vacuum tunneling from the $\phi$-vacuum to the $h$-vacuum. We assume the second stage phase transition from $\phi$- to $h$-vacuum to occur instantaneously without large reheating. 

The Boltzmann equations in the $Z_2$ symmetry phase and the $h$-vacuum are given by
\begin{eqnarray}
s H z \frac{d Y_{\chi}}{d z} &=& - \langle \sigma v \rangle_{\chi \chi \rightarrow N N} Y_{\chi eq}^2 s^2 \left( \frac{Y_{\chi}^2}{Y_{\chi eq}^2} - \frac{Y_N^2}{Y_{N eq}^2} \right) - \langle \sigma v \rangle_{\chi \chi \rightarrow \phi \phi} Y_{\chi eq}^2 s^2 \left( \frac{Y_{\chi}^2}{Y_{\chi eq}^2} - \frac{Y_{\phi}^2}{Y_{\phi eq}^2} \right) \nonumber \\
&-& \langle \sigma v \rangle_{\chi \phi \rightarrow \text{allSM}} s^2 (Y_{\chi} Y_{\phi} - Y_{\chi eq} Y_{\phi eq}) - \bar{\Gamma}_{\phi \rightarrow \chi N} Y_{\phi eq} s \left( \frac{Y_{\chi} Y_{N}}{Y_{\chi eq} Y_{N eq}} - \frac{Y_{\phi}}{Y_{\phi eq}} \right) \nonumber \\
&-&\langle \sigma v \rangle_{\chi \phi \rightarrow N h} s^2 Y_{\chi eq} Y_{\phi eq} \left( \frac{Y_{\chi} Y_{\phi}}{Y_{\chi eq} Y_{\phi eq}}-\frac{Y_N}{Y_{N eq}} \right) - \langle \sigma v \rangle_{\chi N \rightarrow \phi h} s^2 Y_{\chi eq} Y_{N eq} \left( \frac{Y_{\chi} Y_{N}}{Y_{\chi eq} Y_{N eq}} - \frac{Y_{\phi}}{Y_{\phi eq}} \right) \nonumber \\
&-& \langle \sigma v \rangle_{\chi h \rightarrow \phi N} s^2 Y_{\chi eq} Y_{h eq} \left( \frac{Y_{\chi}}{Y_{\chi eq}} - \frac{Y_{\phi} Y_N}{Y_{\phi eq} Y_{N eq}} \right), \nonumber \\
s H z \frac{d Y_{\phi}}{d z} &=& -\langle \sigma v \rangle_{\phi \phi \rightarrow N N} Y_{\phi eq}^2 s^2 \left( \frac{Y_{\phi}^2}{Y_{\phi eq}^2} - \frac{Y_N^2}{Y_{N eq}^2} \right) - \langle \sigma v \rangle_{\phi \phi \rightarrow \chi \chi} Y_{\phi eq}^2 s^2 \left( \frac{Y_{\phi}^2}{Y_{\phi eq}^2} - \frac{Y_{\chi}^2}{Y_{\chi eq}^2} \right) \nonumber \\
&-& \langle \sigma v \rangle_{\phi \phi \rightarrow \text{allSM}} s^2 ( Y_{\phi}^2 - Y_{\phi eq}^2 ) - \langle \sigma v \rangle_{\chi \phi \rightarrow \text{allSM}} s^2 (Y_{\chi} Y_{\phi} - Y_{\chi eq} Y_{\phi eq}) \nonumber \\
&-& \bar{\Gamma}_{\phi \rightarrow \chi N} Y_{\phi eq} s \left( \frac{Y_{\phi}}{Y_{\phi eq}} - \frac{Y_{\chi} Y_{N}}{Y_{\chi eq} Y_{N eq}} \right) - \langle \sigma v \rangle_{\chi \phi \rightarrow N h} s^2 Y_{\chi eq} Y_{\phi eq} \left( \frac{Y_{\chi} Y_{\phi}}{Y_{\chi eq} Y_{\phi eq}} - \frac{Y_N}{Y_{N eq}} \right) \nonumber \\
&-& \langle \sigma v \rangle_{\phi h \rightarrow \chi N} s^2 Y_{\phi eq} Y_{h eq} \left(\frac{Y_{\phi}}{Y_{\phi eq}} - \frac{Y_{\chi} Y_N}{Y_{\chi eq} Y_{N eq}} \right) - \langle \sigma v \rangle_{\phi N \rightarrow \chi h} s^2 Y_{\phi eq} Y_{N eq} \left( \frac{Y_{\phi} Y_{N}}{Y_{\phi eq} Y_{N eq}} - \frac{Y_{\chi}}{Y_{\chi eq}} \right), \nonumber \\
s H z \frac{d Y_N}{d z} &=& -\langle \sigma v \rangle_{N N \rightarrow \chi \chi} Y_{N eq}^2 s^2 \left( \frac{Y_N^2}{Y_{N eq}^2} - \frac{Y_{\chi}^2}{Y_{\chi eq}^2} \right) - 2 \langle \sigma v \rangle_{N N \rightarrow \phi \phi} Y_{N eq}^2 s^2 \left( \frac{Y_N^2}{Y_{N eq}^2} - \frac{Y_{\phi}^2}{Y_{\phi eq}^2} \right) \nonumber \\
&-& \bar{\Gamma}_{N} s (Y_N - Y_{N eq}) - \langle \sigma v \rangle_{N h \rightarrow \chi \phi} s^2 Y_{N eq} Y_{h eq} \left( \frac{Y_N}{Y_{N eq}} - \frac{Y_{\chi} Y_{\phi}}{Y_{\chi eq} Y_{\phi eq}} \right)  \nonumber \\
&-& \langle \sigma v \rangle_{\chi N \rightarrow \phi h} s^2 Y_{\chi eq} Y_{N eq} \left( \frac{Y_{\chi} Y_{N}}{Y_{\chi eq} Y_{N eq}} - \frac{Y_{\phi}}{Y_{\phi eq}} \right) - \langle \sigma v \rangle_{\phi N \rightarrow \chi h}  s^2 \left( \frac{Y_{\phi} Y_N}{Y_{\phi eq} Y_{N eq}} - \frac{Y_{\chi}}{Y_{\chi eq}} \right) \nonumber \\
&+& \bar{\Gamma}_{\phi \rightarrow \chi N} Y_{\phi eq} s \left(  \frac{Y_{\phi}}{Y_{\phi eq}} - \frac{Y_{\chi} Y_{N}}{Y_{\chi eq} Y_{N eq}} \right) - \langle \sigma v \rangle_{N N \rightarrow \text{allSM}} s^2 (Y_N^2 - Y_{N eq} ^2)
, \label{Boltzhphase}
\end{eqnarray}
where the $Y_X= \frac{n_X}{s}$ is the actual number of the constituent $X$ per-comoving-volume, and the $Y_{X eq} = \frac{n_{X eq}}{s}$ is the equilibrium number of the constituent $X$ per-comoving-volume, $n_{X (eq)}$ is the (equilibrium) number density of the constituent $X$, s is the entropy density, $z=\frac{m_\chi}{T}$, and $T$ is the temperature, $H$ is the Hubble constant. Here, we note that the reheating temperature is supposed to be above the $\phi$-vacuum to $h$-vacuum phase transition temperature, which is taken to be the bubble nucleation temperature as been estimated in Fig.~\ref{fig:GW} of Sec.~\ref{sec:GWTn}.  

Firstly, we define the thermal decay width of $\bar{\Gamma}_{\phi \rightarrow \chi N}$ as
\begin{eqnarray}
\bar{\Gamma}_{\phi \rightarrow \chi N} = \frac{K_1(\frac{m_\phi(T)}{T})}{K_2(\frac{m_\phi(T)}{T})} \Gamma_{\phi \rightarrow \chi N}\;, \label{eq:kin:phi-chi-N1} 
\end{eqnarray}
with the thermal mass $m_{\phi}(T)$ calculated as in the previous section and 
\begin{eqnarray}
  \Gamma(\phi \to \chi \bar N)
   & =& \frac{y_{\chi}^2}{8 \pi}
       \frac{m_\phi^2(T) - (m_\chi + m_N)^2}{m_\phi^3(T)}\nonumber\\
				&&\times
       \sqrt{ \big(m_\phi^2(T) - (m_\chi + m_N)^2 \big)
              \big( m_\phi^2(T) - (m_\chi - m_N)^2 \big) } \;.
  \label{eq:kin:phi-chi-N2} 
\end{eqnarray} 
If one have $m_\chi>m_\phi(T)+m_N$ temporary during the phase transition process, we need 
 need to replace the $\Gamma(\phi \to \chi \bar N)$ by the decay width of the $\chi$, i.e., $\Gamma(\chi\to N \phi)$, being given by

\begin{eqnarray}
    \Gamma(\chi\to N \phi)
    &=& \frac{y_{\chi}^2}{16 \pi}
       \frac{m_\chi^2-(m_\phi(T)+ m_N)^2 }{m_\chi^3}\nonumber\\
				&&
				\times
       \sqrt{ \big( m_\chi^2 - (m_N+ m_\phi(T))^2 \big)
              \big(m_\chi^2 - (m_N - m_\phi(T))^2 \big) } \,.
  \label{eq:kin:N-l-h}  
\end{eqnarray}

For completeness, we consider also the decay (or inverse-decay processes) $N \rightarrow h^{\pm 0} l$( or $h^{\pm 0} \rightarrow l N$). 
 After the phase transition of $\phi$- to $h$-vacuum, we need to consider the processes $N \leftrightarrow W^{\pm}/Z/h$ also. As in Ref.~\cite{Liumangfa}, we calculate the $N \rightarrow h^{\pm 0} l$ or $h^{\pm 0} \rightarrow l N$ with all the four states of the Higgs doublets having the Higgs boson mass $m_h(T)$. If $m_N > m_h(T)$,
\begin{eqnarray}
\tilde{\Gamma}_N = \frac{K_1 (\frac{m_N}{T})}{K_2 (\frac{m_N}{T})} \Gamma_{N \rightarrow H + l}\;,
\end{eqnarray}
while $m_N < m_h(T)$,
\begin{eqnarray}
\tilde{\Gamma}_N = \frac{Y_{H eq}}{Y_{N eq}} \frac{K_1 (\frac{m_h(T)}{T})}{K_2 (\frac{m_h(T)}{T})} \; \Gamma_{H \rightarrow N l}. \label{InverseNDecay}
\end{eqnarray}
As the temperature cools down to $T \ll m_h(= 125 \text{ GeV})$, the magnitude of the $\tilde{\Gamma}_N$ is highly suppressed by a factor of $e^{\frac{-2 m_h + m_N}{T}}$. In order to let the right-handed neutrino decay, we use
\begin{eqnarray}
\tilde{\Gamma}_N = \frac{K_1 (\frac{m_N}{T})}{K_2 (\frac{m_N}{T})} \Gamma_{N \rightarrow h^*/W^*/Z^* l}\;,
\end{eqnarray}
when the $\tilde{\Gamma}_N< \frac{K_1 (\frac{m_N}{T})}{K_2 (\frac{m_N}{T})} \Gamma_{N \rightarrow h^*/W^*/Z^* l}$ with $\Gamma_{N \rightarrow h^*/W^*/Z^* l}$ being calculated at the zero temperature.  We further note that in this work, the thermal history of the sterile neutrino does not significantly affect the dark matter production.
In this work, we restrict our interest to the mass of $N$ being smaller than 80 GeV. In this case, these two body decays are kinematically forbidden.  Meanwhile,
three body decays is dominant by the process involving the exchange of a virtual $W$~\cite{Babu:2014uoa}, 
\begin{equation}
\Gamma(N \rightarrow 3~{\rm body}) = \frac{G_F^2 M_N^5}{192 \pi^3} \sin^2\theta_{\nu N}\left(1 + \frac{3}{5} \frac{M_N^2}{m_W^2}\right) (2) \left[5 + 3 F\left(\frac{m_c^2}{M_N^2}\right) + F\left(\frac{m_\tau^2}{M_N^2}\right) \right]~.
\label{3body}
\end{equation}
with the kinematic function $F(x) = \{1- 8 x + 8 x^3 - x^4-12 x^2\, {\rm ln} x\}$.  An overall factor of 2 appears in the bracket for the $N$ being Majorana case. This factor does not appear in the Dirac case due to the interference of the two components decay. As for the pseudo-Dirac particles, the splitting of the two components will remove the interference, leaving a factor between 1 and 2. In this paper, we adopt the factor 2, but practically, the detailed value does not affect the qualitative, and most of the quantitative results. 

Secondly, the thermally averaged cross section times velocity
$\langle \sigma v \rangle_{AB \rightarrow CD}$ is given by
\begin{eqnarray}
\langle \sigma v \rangle_{A B \rightarrow C D} = \frac{1}{(1+\delta_{CD}) n_A n_B} \frac{g_A g_B T}{32 \pi^4} \int ds^{\prime} s^{\prime \frac{3}{2}} K_1 \left( \frac{\sqrt{s^{\prime}}}{T} \right) \lambda \left( 1, \frac{m_A^2}{s^{\prime}}, \frac{m_B^2}{s^{\prime}} \right) \sigma_{A B \rightarrow C D} (s^{\prime}),
\end{eqnarray}
where $\delta_{CD} = 1(0)$ if $C$ and $D$ are identical(different) particles, $g_A$ and $g_B$ are the degrees of freedoms of particle $A$ and $B$, $K_1$ is a Bessel function, $\sigma_{A B \rightarrow C D}(s^{\prime})$ is the cross section of the process $A B \rightarrow C D$ with the total energy in the center of mass frame is $\sqrt{s^{\prime}}$. We note that all thermal effects being explored in Sec.~\ref{sec:mvevT} has been implemented here.

On the other hand, considering the reheating temperature to above the temperate of the first-stage second-order phase transition, there should be other terms arising from the breaking of the $Z_2$ symmetry,
\begin{eqnarray}
s H z \frac{d Y_{\chi}}{d z} &+=& -\langle \sigma v \rangle_{\phi \chi \rightarrow \phi N} s^2 \left( Y_{\phi} Y_{\chi} - Y_{\phi} Y_{N} \frac{Y_{\chi eq}}{Y_{N eq}} \right) - \langle \sigma v \rangle_{\chi N \rightarrow \phi \phi} \left( Y_{\chi} Y_N - \frac{Y_{\chi eq} Y_{N eq}}{Y_{\phi eq} Y_{\phi eq}} Y_{\phi} Y_{\phi} \right), \nonumber \\
& & - 4 \langle \sigma v \rangle_{h \chi \rightarrow h N} Y_{h eq} Y_{\chi eq} \left( \frac{Y_{\chi}}{Y_{\chi eq}} - \frac{Y_{N}}{Y_{N eq}} \right), \nonumber \\
s H z \frac{d Y_{\phi}}{d z} &+=& -\langle \sigma v \rangle_{\phi \phi \rightarrow \chi N} \left( Y_{\phi} Y_{\phi} - \frac{Y_{\phi eq} Y_{\phi eq}}{Y_{\chi eq} Y_{N eq}} Y_{\chi} Y_{N} \right), \nonumber \\
s H z \frac{d Y_N}{d z} &+=& -\langle \sigma v \rangle_{\phi N \rightarrow \phi \chi} \left( Y_{\phi} Y_{N} - \frac{Y_{N eq}}{Y_{\chi eq}} Y_{\phi} Y_{\chi} \right) - \langle \sigma v \rangle_{\chi N \rightarrow \phi \phi} \left( Y_{\chi} Y_N - \frac{Y_{\chi eq} Y_{N eq}}{Y_{\phi eq} Y_{\phi eq}} Y_{\phi} Y_{\phi} \right) \nonumber \\
& &-4 \langle \sigma v \rangle_{h N \rightarrow h \chi} Y_{h eq} Y_{\chi eq} \left( \frac{Y_{\chi}}{Y_{\chi eq}} - \frac{Y_{N}}{Y_{N eq}} \right).
 \label{Boltzmann_Supplement}
\end{eqnarray}
The number ``4'' appearing in the $\langle \sigma v \rangle_{h \chi \leftrightarrow h N}$ indicates that in the $\phi$-vacuum, all the charged and neutral four Higgs bosons are identical. We should note that an on-shell t/u-channel particle appears in many of the processes in the (\ref{Boltzmann_Supplement}), due to the on-shell decay of the initial particles. Therefore, when $m_{\chi}>m_{N}+m_{\phi}$ or $m_{\phi}>m_{N}+m_{\chi}$, we do not include all the terms in the (\ref{Boltzmann_Supplement}). And that all terms involving the h$\phi\phi$ vertices should be take into account since the EW symmetry is respected in the $\phi-$vacuum.

To solve the Boltzmann equations of Eq.~(\ref{Boltzhphase}, \ref{Boltzmann_Supplement}), we use the ready-made function \cite{Stiff1, Stiff2} embeded in the micrOMEGAs \cite{micrOMEGAs} for computing the stiff equations Eqn.~(\ref{Boltzhphase}, \ref{Boltzmann_Supplement}). With the model file being prepared by FeynRules \cite{FeynRules}, after implemented the thermal masses and VEVs of $\phi$ and $h$
as a function of temperature as in Sec.~\ref{sec:mvevT},
we calculate all the thermal $\langle \sigma v \rangle (s)$ and the thermal decay widths of the particles using the CalcHEP \cite{CalcHEP} embeded in the micrOMEGAs.  Since the first step phase transition to the $\phi$-vacuum is second order, one can expect negligible entropy injection, the strongly first order phase transition of $\phi$-vacuum to $h$-vacuum injected entropy is estimated to be around precent level in Sec.~\ref{sec:mvevT}. The degree of freedoms $g_*$ and $g_{*S}$ implemented in the micrOMEGAs are adopted in order to calculate the Hubble constant $H = 1.66 \sqrt{g_*} \frac{T^2}{M_{pl}}$, and entropy $s = \frac{2 \pi^2}{45} g_{*S} T^3$. Here the Planck scale is $M_{pl} = 1.22 \times 10^{19} \text{ GeV}$.

\subsection{On the decouple conditions bounds on the parameter spaces}

\begin{figure}[!htp]
\begin{center}
\includegraphics[width=0.6\textwidth]{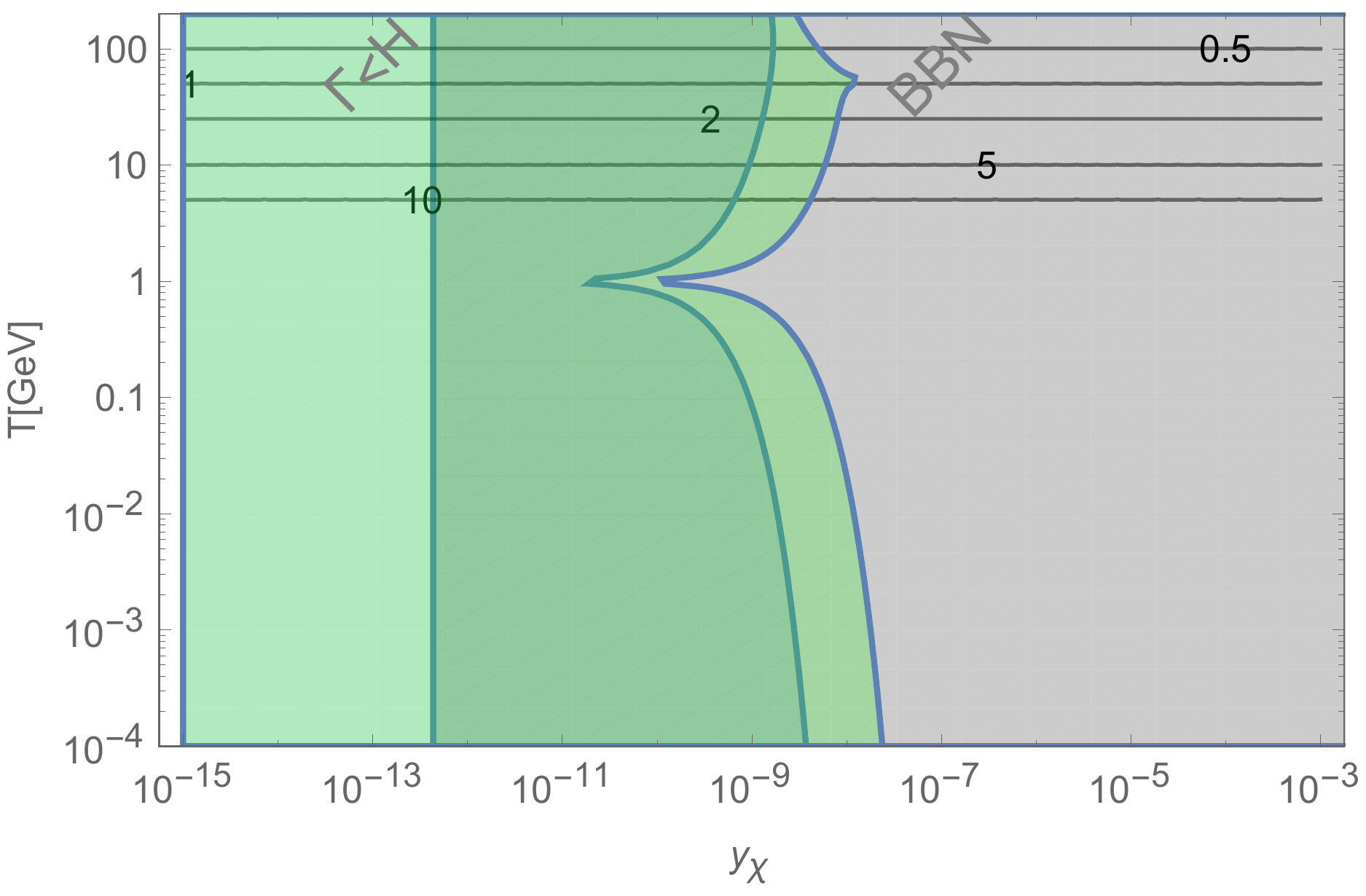}
\caption{Decouple conditions bounds on the yukawa of $y_\chi$, the values of $z=m_\chi/T$ are shown with solid lines. The Blue and Green regions corresponds to the $m_N=20$ GeV, $m_\phi=110 $ GeV, $m_\chi=50$ GeV, ($m_\phi>m_\chi+m_N$) and $m_\chi=180$($m_\chi>m_\phi+m_N$) respectively. The Gray region are BBN required on the coupling of $y_\chi$.}
\label{fig:dec}
\end{center}
\end{figure}

For the FIMP production of $\chi$, that would never reach equilibrium, which can be used to 
estimate the bounds on the couplings through the decouple condition $\Gamma<H$.
For the decay or inverse induced FIMP, we just needs $\Gamma_{\chi }<H$,
with decay width being given by Eq.~\ref{eq:kin:phi-chi-N1} and Eq.~\ref{eq:kin:phi-chi-N2}. Which results in the bounds on the $y_\chi$ as shown in Fig.~\ref{fig:dec}. The Figure illustrates that for FIMP production process, the typical  freeze in temperature $T_{fi}$($\sim m_\chi/z$ with $z\sim 1$) will be smaller than  
the phase transition temperature $T_n$. Therefore, one
can expect the phase transition process will affect the FIMP process. 

\subsection{ Dark matter }

\begin{figure}
\includegraphics[width=0.6\textwidth]{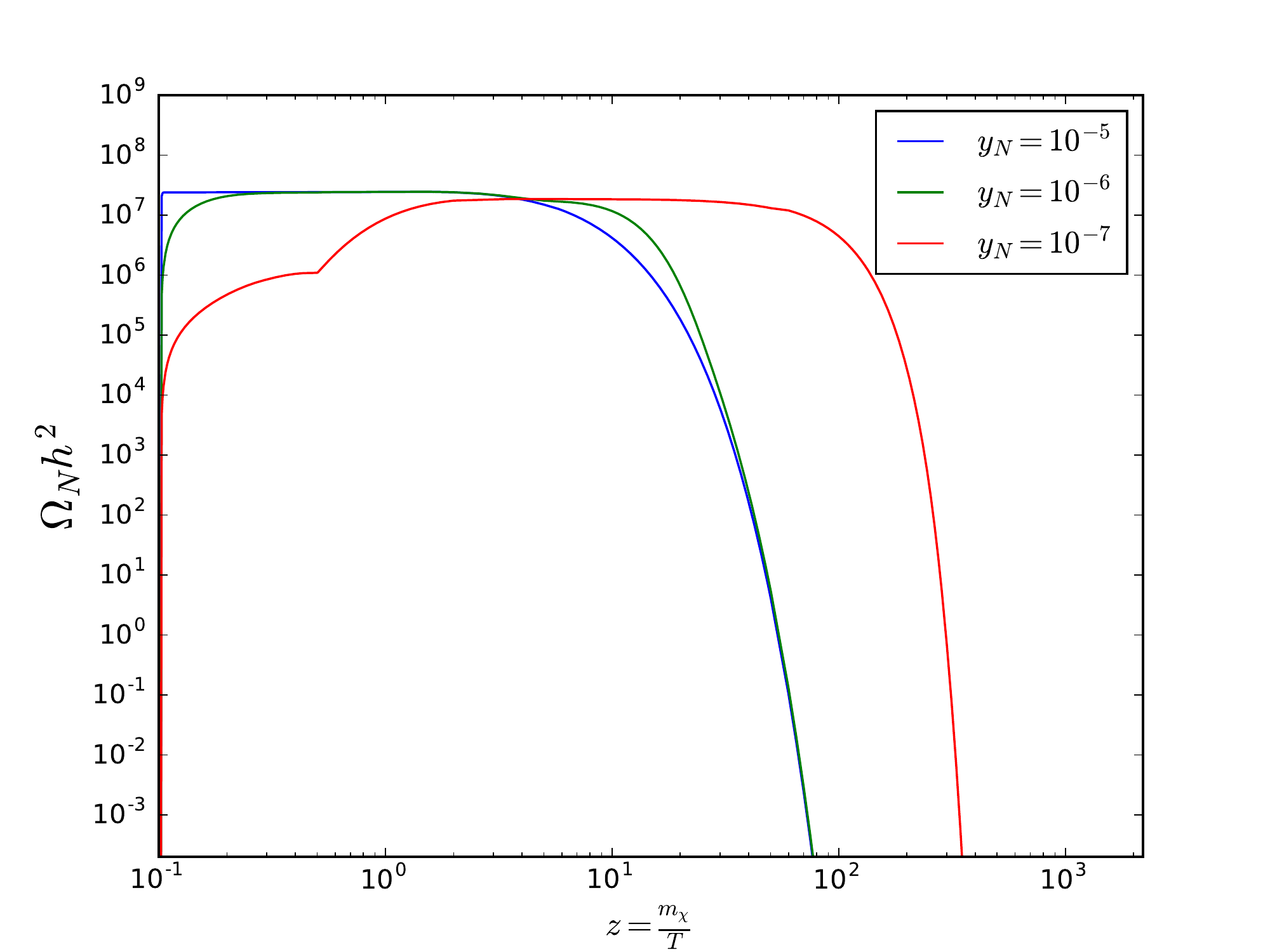}
\caption{The thermal evolution of the sterile neutrino, in the case of three different $y_N$. We should note that the $y_N=10^{-5}$ curve is very close to the equilibrium case, besides the very sharp ``jumping up'' at the beginning.} \label{NThermal}
\end{figure}

Firstly, we note that the sterile neutrino can become in thermal equilibrium with the thermal bath through its decay/inverse decay shortly after the reheating process for the typical $y_N>10^{-6}$. 
And we note that the values of the $y_N$ does not have a strong impact on the thermal history of the dark matter. However, since there are only two couplings $y_{\chi}$ and $y_N$ which connect the sterile neutrino with the SM sector, and usually $y_{\chi} \ll y_N$, the thermal history of the sterile neutrino is mainly decided by $y_N$. We show in the Fig.~\ref{NThermal} the thermal evolutions of $y_N=10^{-5}$, $10^{-6}$, $10^{-7}$ respectively. We can see that $10^{-6}$ is some critical value, below which the sterile neutrino evolution will be similar to a ``freeze-in'' process, and its decay will be delayed significantly.

\begin{figure}[!htp]
\begin{center}
\includegraphics[width=0.4\textwidth]{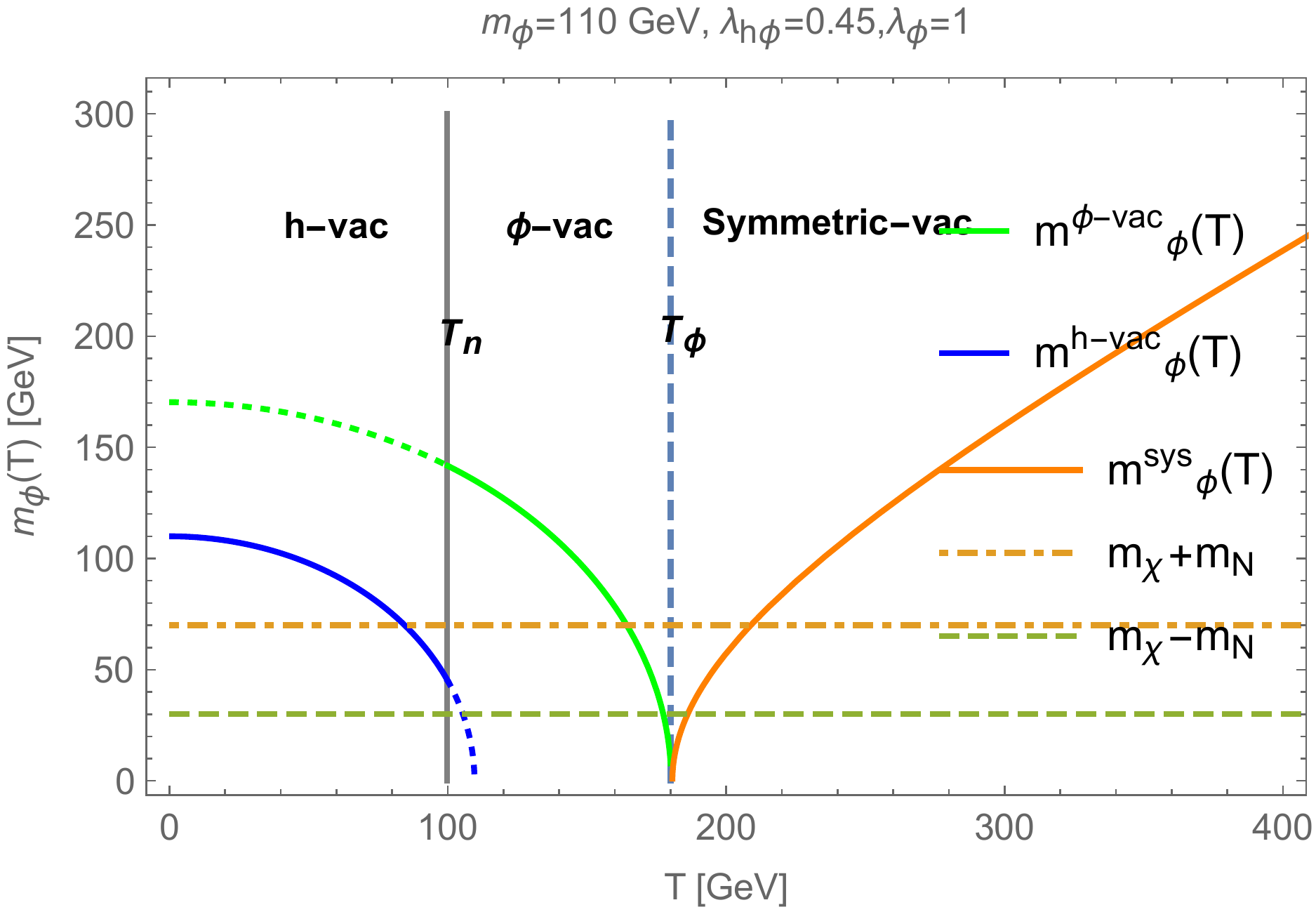}
\includegraphics[width=0.4\textwidth]{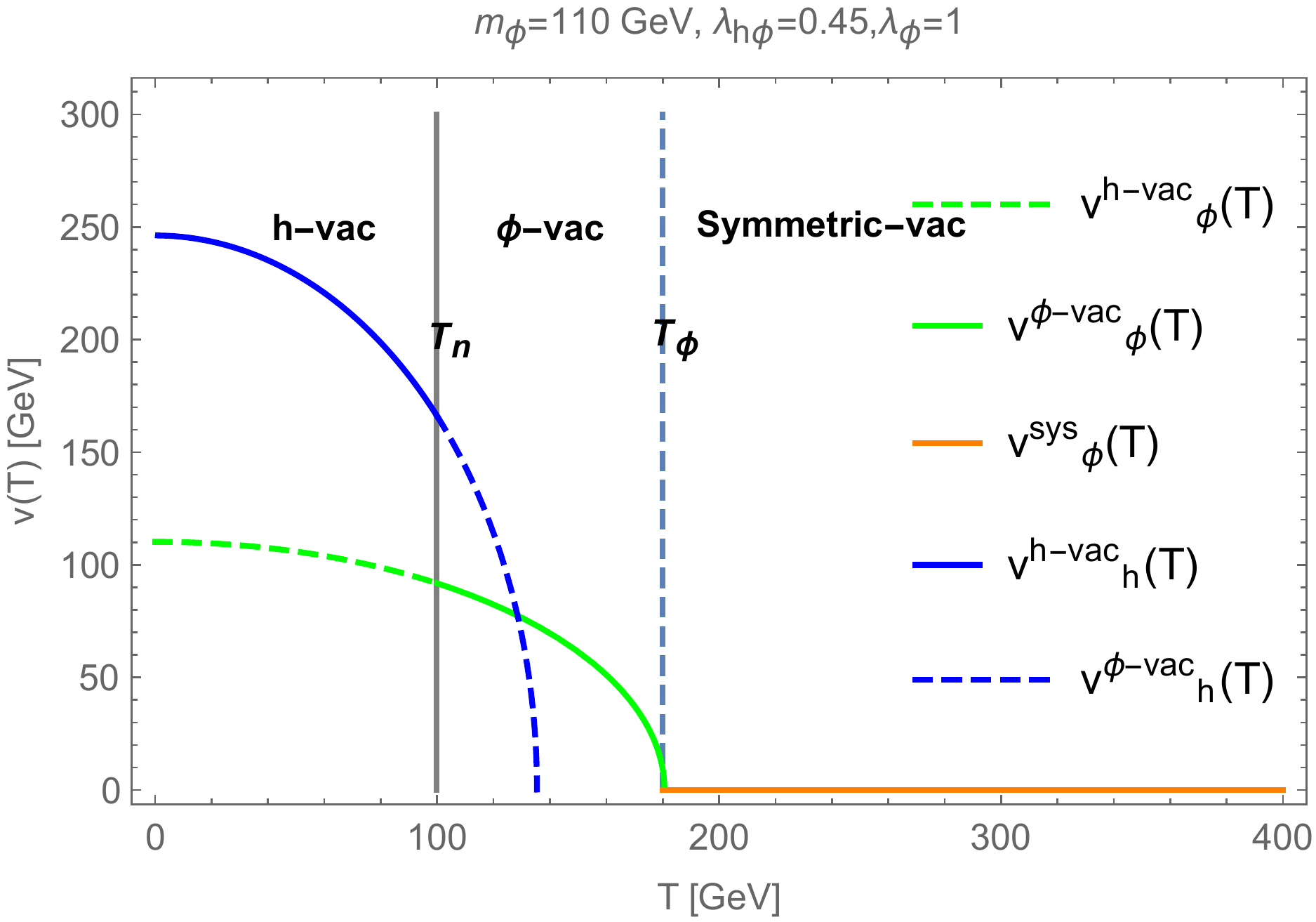}
\includegraphics[width=0.4\textwidth]{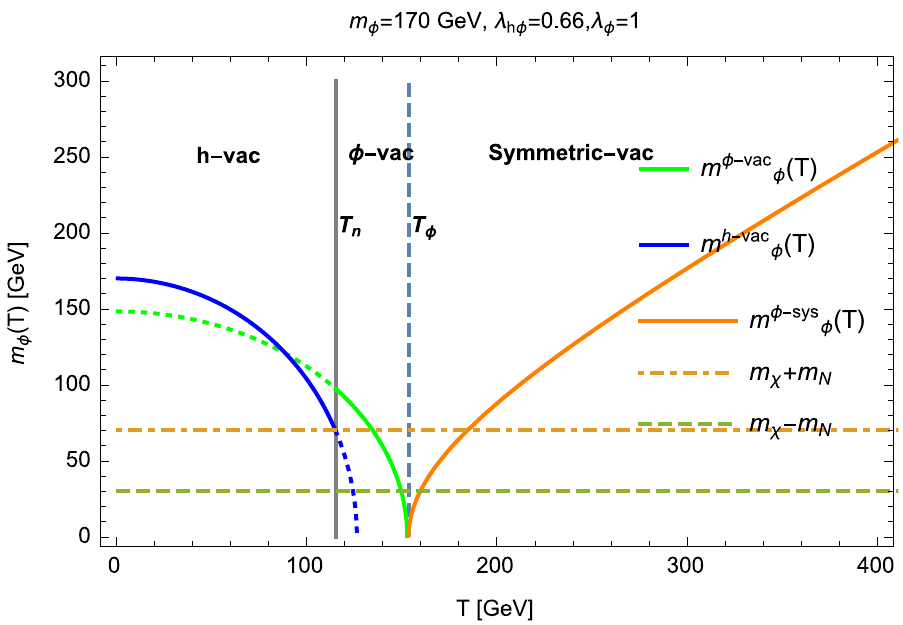}
\includegraphics[width=0.4\textwidth]{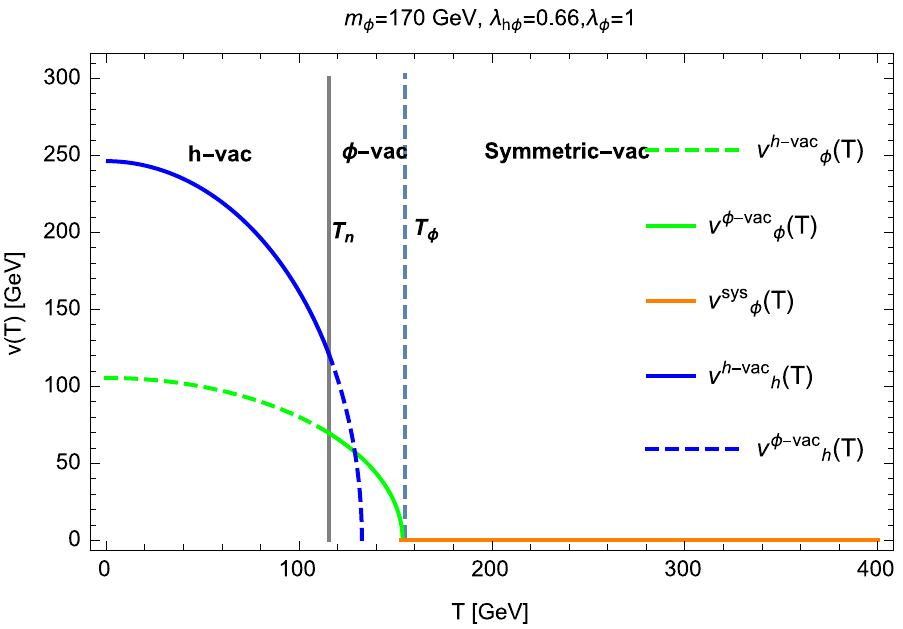}
\caption{The thermal corrected mass evolution of $\phi$ and the evolution of the VEVs versus temperature (T) for $m_\phi=110$ GeV (to be studied in Fig.~\ref{fig:BPDMphi} and the top-panel of Fig.~\ref{fig:FIMPchi}), $m_\phi=170$ GeV (to be studied in Fig.~\ref{fig:FIMPchiBP3}).}
\label{fig:FIMPmphi}
\end{center}
\end{figure}

Since the SFOEWPT prefer a moderate $\lambda_{h\phi}$, with which the $\phi$ particle can be produced by freeze out mechanism, we study $\chi$ freeze in scenario to reveal the EWPT effects. 
Before the study of DM production details, we explore the mass threshold related to the following DM analysis as the temperature cools down, see Fig.~\ref{fig:FIMPmphi}. Before the $Z_2$ respected by $\phi$ and EW symmetry is broken, we denote the symmetric phase in the figure, the $m_\phi(T)$ is dominated by the thermal corrections at temperatures higher than the second order phase transition temperature of $Z_2$ (we denote $T_\phi$) where there is no VEV for $h$ or $\phi$ field. In this symmetric phase, one have  $m_\phi(T)=m_\phi^{sys}(T)$ and $v_{h,\phi}=0$. During the temperature of $T_\phi$ and the strongly first order phase transition from the $Z_2$ broken EW symmetry phase ($\phi-$vacuum) to the $Z_2$ preserved EW broken phase ($h$-vacuum), we have the finite temperature mass of $m_\phi(T)=m_\phi^{\phi-vac}(T)$ in the $\phi$-vacuum with the accompanied VEV of $\phi$ being $v_\phi^{\phi-vac}(T)$. As the temperature cools down further to $T_n$, the Universe go through the broken of EW symmetry and restoration of the $Z_2$ symmetry respected by the $\phi$ field. After which, we have $m_\phi(T)=m_\phi^{h-vac}(T)$ in the $h$-vacuum with the VEV of $h$ being $v_h^{h-vac}(T)$. With the Universe further cools down, we finally locate in the vacuum of $U(1)_{em}$ symmetry with the $v_h^{h-vac}(T=0)=246 $GeV. Our study shows that, a larger quartic coupling of $\lambda_{h\phi}$ and a larger $m_\phi$ will leads to a smaller $T_\phi$, means a later happening of the first-stage second-order phase transition.  

\subsubsection{$\phi-$DM}

In this section, we consider $m_\phi+m_{N}<m_\chi$, in this case we have $\phi$ as dark matter candidate. When one consider the $\phi$ produced from the FIMP mechanism as in~Ref.\cite{Yaguna:2011qn} the SFOEWPT fails. 
The relic abundance of $\phi$ can be obtained from the late decay of $ \chi \to\phi N $ after $\chi$ is generated, while the freeze out mechanism can also contribute significantly when 
the $\phi$ couples with the SM Higgs moderately for a successfull SFOEWPT.
One can simplify the evaluation of the relic density of the $\phi$ from the two contributions, one from freeze-out and the other from the late decay of $\chi$. 
\begin{equation}
\Omega_{\phi}\,h^{2}=\Omega_{\phi}^{freeze-out}\,h^{2}+\Omega_{\phi}^{\chi-decay}\,h^{2}
\end{equation}
with 
\begin{equation}
\Omega_{\phi}^{\chi-decay}\,h^{2}=\frac{m_{\chi}}{m_\phi}\,\Omega_{\chi}^{freeze-in}\,h^{2}.
\end{equation}

\begin{figure}[!ht]
\begin{center}
\includegraphics[width=0.3\textwidth]{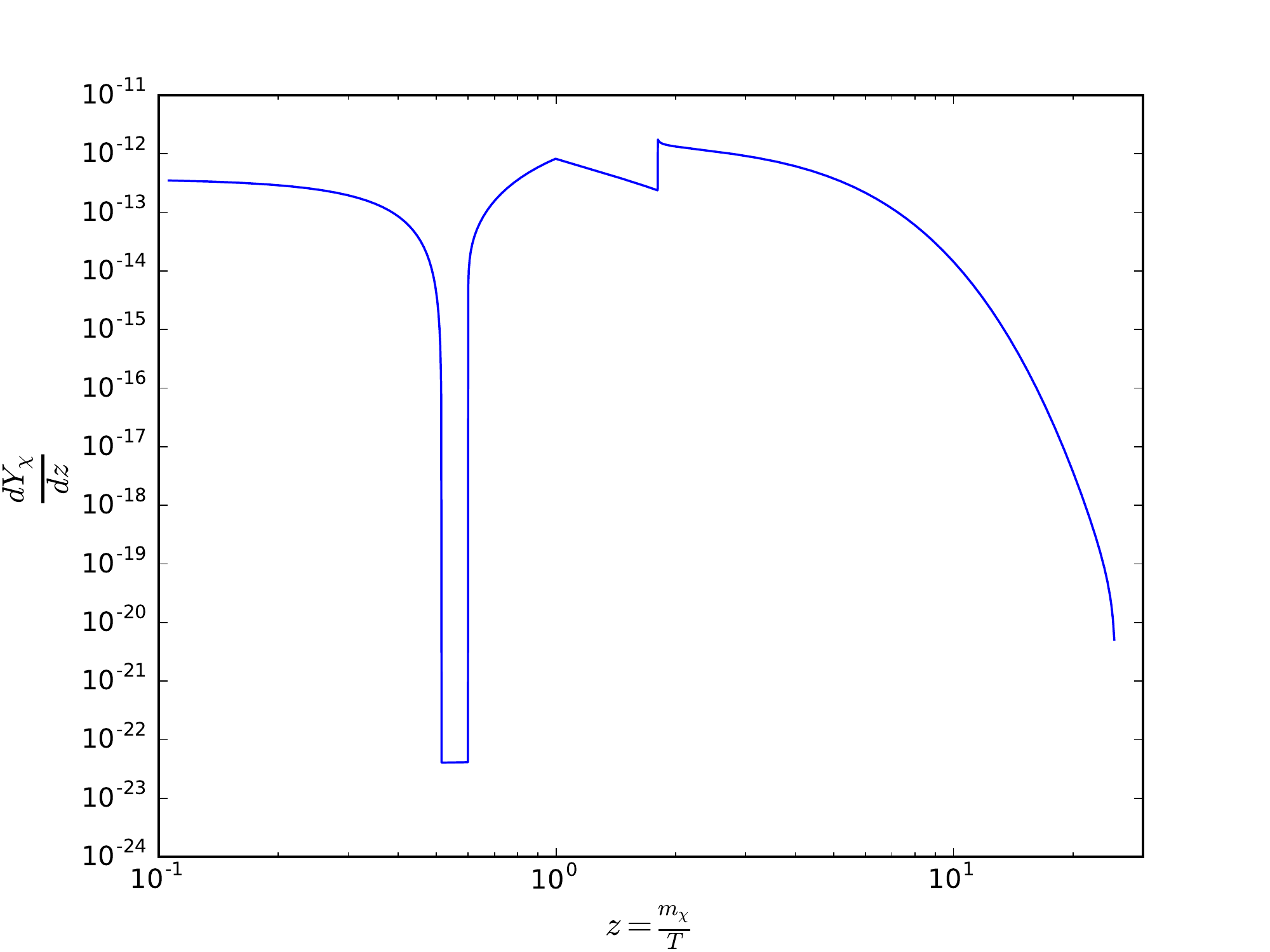}
\includegraphics[width=0.3\textwidth]{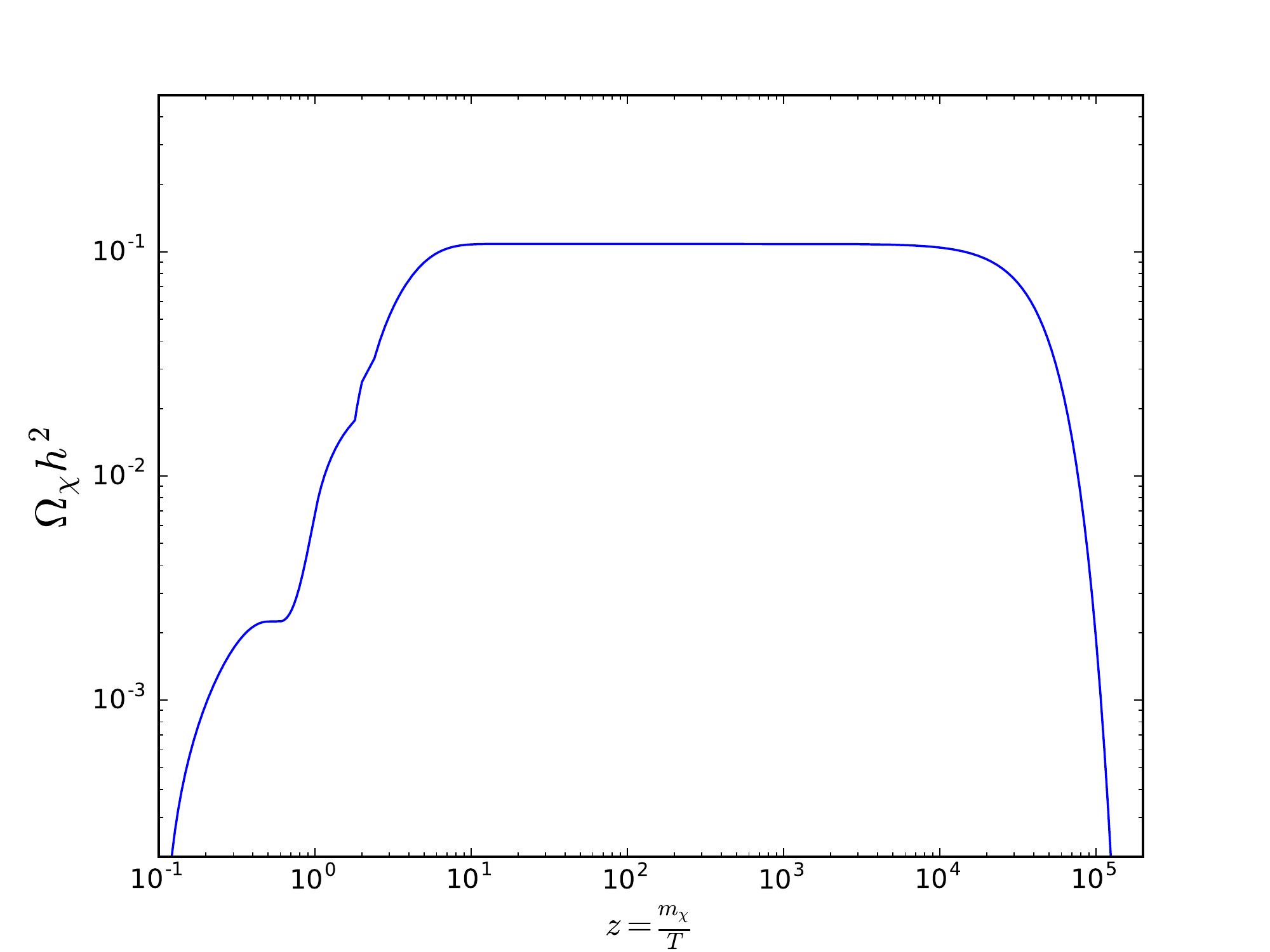}
\includegraphics[width=0.3\textwidth]{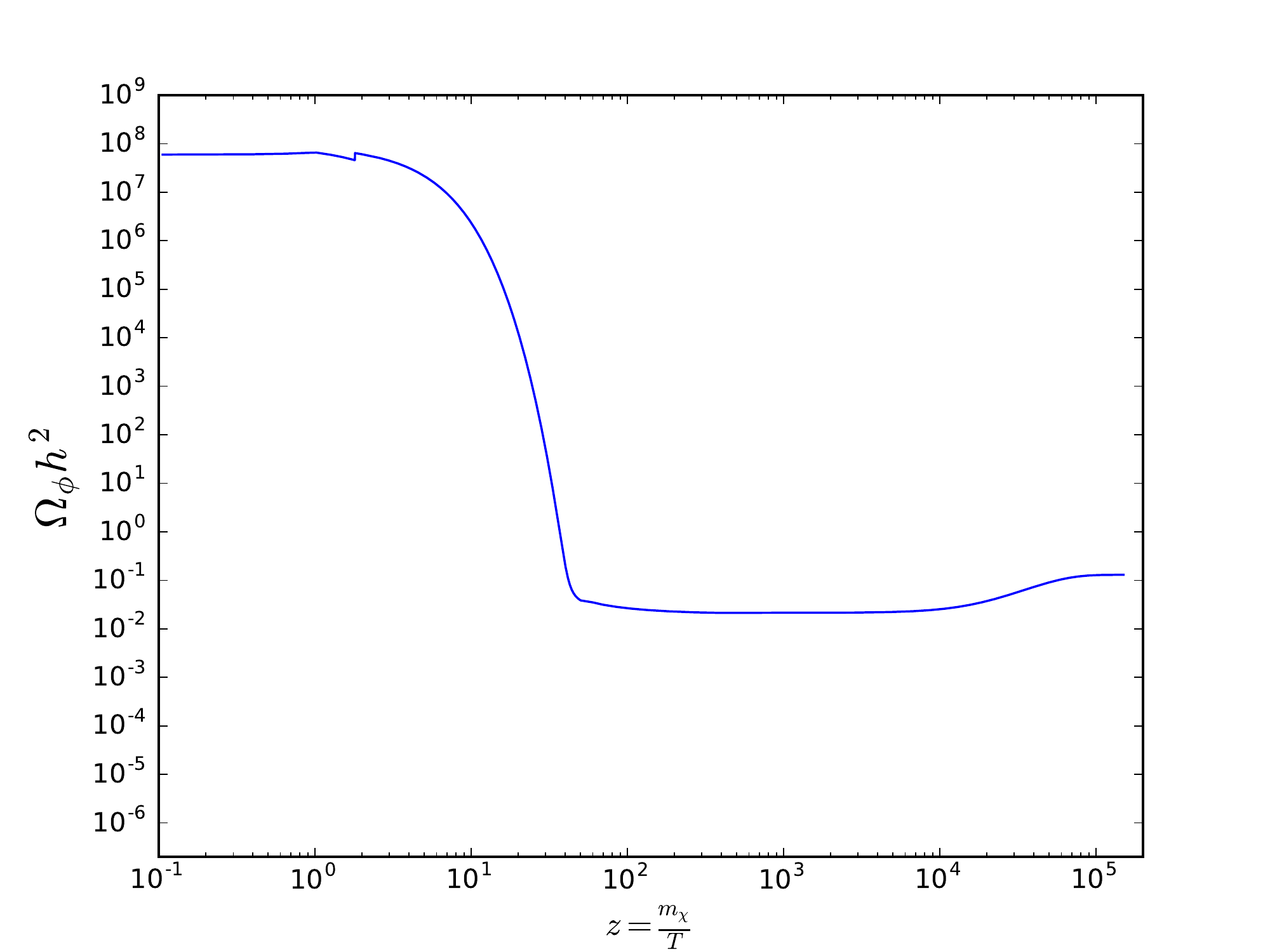}
\caption{The evolution of the number densitys of $\chi$ and $\phi$ for $m_\phi<m_\chi$ with $m_\phi=110$ GeV, $\lambda_{h\phi}(\lambda_\phi)=0.45(1)$, $m_\chi=180$ GeV, and $m_N=20$ GeV.}
\label{fig:BPDMphi}
\end{center}
\end{figure}

Note that the productions of the dark matter can be separated into different stages, which is similar, but much simpler than the stages in the next subsection. For our benchmark point, since the $m_{\chi}$ is much larger, after the first-stage second-order phase-transition, the $m_{\phi}$ never exceeds the $m_{\chi}-m_{N}$. Therefore the structure in the Fig.~\ref{fig:BPDMphi} are much simpler. Finally, when the $z \gg 100$, most of the $\chi$ decays to the $\phi$.  
For the reasons to be described below, the direct detection experiments do not favor the $\phi$-dark matter case. Therefore, we do not illustrate the detailed steps in this subsection, and leave the descriptions in the $\chi$-dark matter case below.

Generally, a small fraction of DM relic abundance from the freeze-out mechanism calls for a larger $\lambda_{h\phi}$ that is needed for a SFOEWPT, 
while a larger $\lambda_{h\phi}$ can easily been excluded by the direct detection experiments of Xenon 1T ~\cite{Aprile:2017iyp,Aprile:2018dbl}, LUX~\cite{Akerib:2016vxi} and Panda X-II ~\cite{Tan:2016zwf} for our interesting DM mass range, see also Ref.~\cite{Cheng:2018ajh}.  On the other hand, a larger fraction of DM relic abundance from the freeze-out requires a relatively small $\lambda_{h\phi}$ which fails the SFOEWPT. This ambiguity almost rules out the possibility of the $\phi$ as dark matter. 

\subsubsection{ $\chi$ as DM}

In this section, we study both the case of $m_\phi>m_\chi+m_N$ and $m_\phi<m_\chi+m_N$, with $\chi$ being DM candidate. 
Eq.(\ref{Boltzhphase},\ref{Boltzmann_Supplement}) tells us that both the $1\leftrightarrow 2$ and the $2\leftrightarrow 2$ channels contribute to the freeze-in processes. Practically, the $\chi \chi \rightarrow N N$ contributions are highly suppressed by the extremely small coupling constants on their extra vertices compared with the $1\leftrightarrow 2$ processes. The dominant $2 \rightarrow 2$ process appears to be $\chi \phi \rightarrow N h$, which makes a comparable contribution with the $1 \leftarrow 2$ processes in the h-vacuum, since its extra vertex is proportional to the $\lambda_{h\phi} v$, which can be relatively large. The $\chi$ freeze-in processes can be affected due to the kinematical threshold can be changed by the thermal effects. We use two typical benchmarks to show the thermal effects in the freeze-in process. The difference is if the $1 \leftrightarrow  2$ processes is active or not when one do not take into account the thermal effects. We show the two scenarios in top and bottom panels of Fig~\ref{fig:FIMPchi}. We show the 
 traditional calculations results for comparison when the thermal effect is ignored, denoted as "No Phase Transition" in the figure. The thermal effects modified scenario are denoted as "With Phase Transition".
We first explain the physical picture of the thermal effects modified scenario shown in the top-left panel (where the  $1 \leftarrow 2$ processes is always active when the thermal effects are not considered, say the  "No Phase Transition" case.) in details:

\begin{figure}[!ht]
\begin{center}
\includegraphics[width=0.3\textwidth]{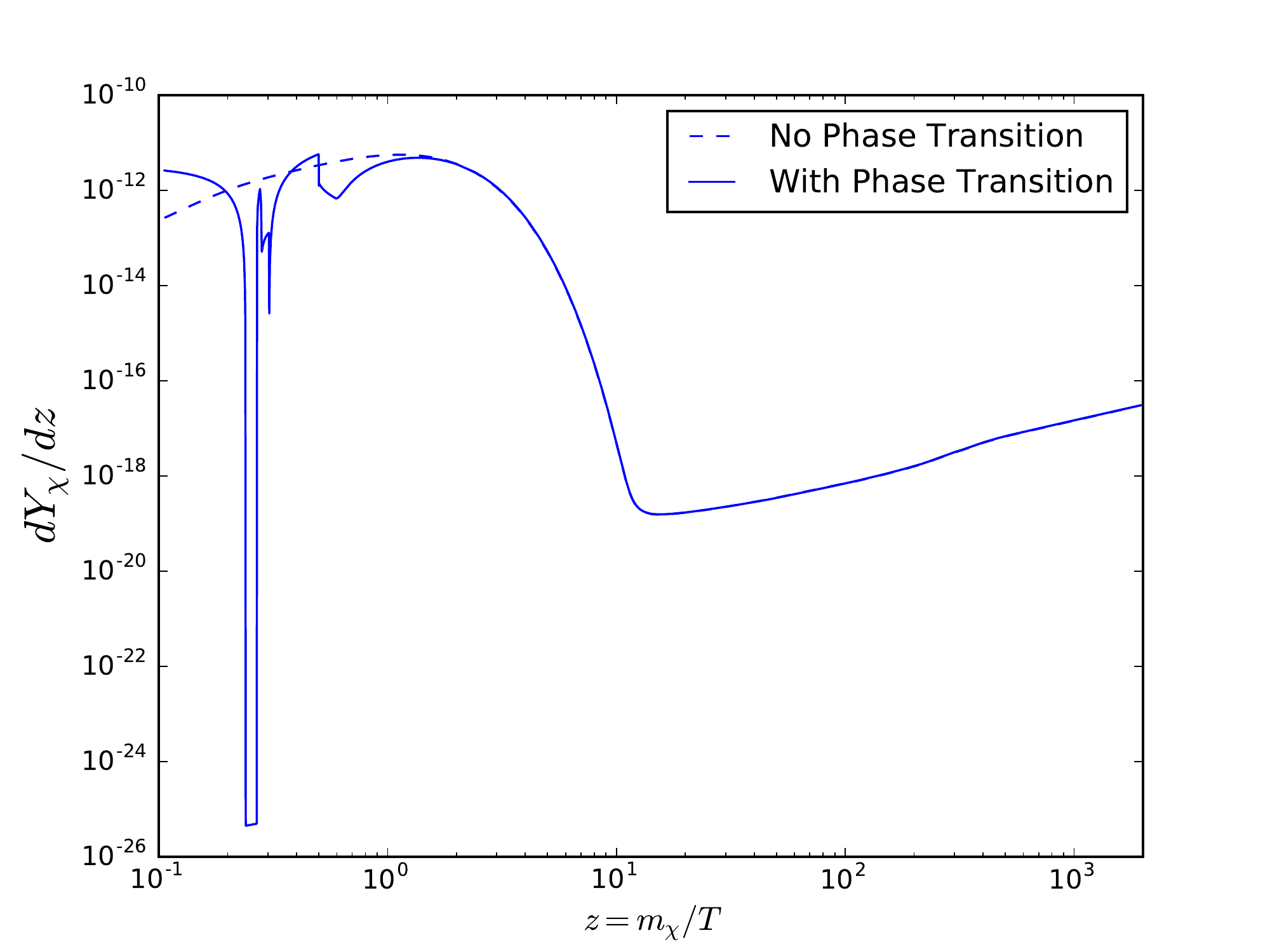}
\includegraphics[width=0.3\textwidth]{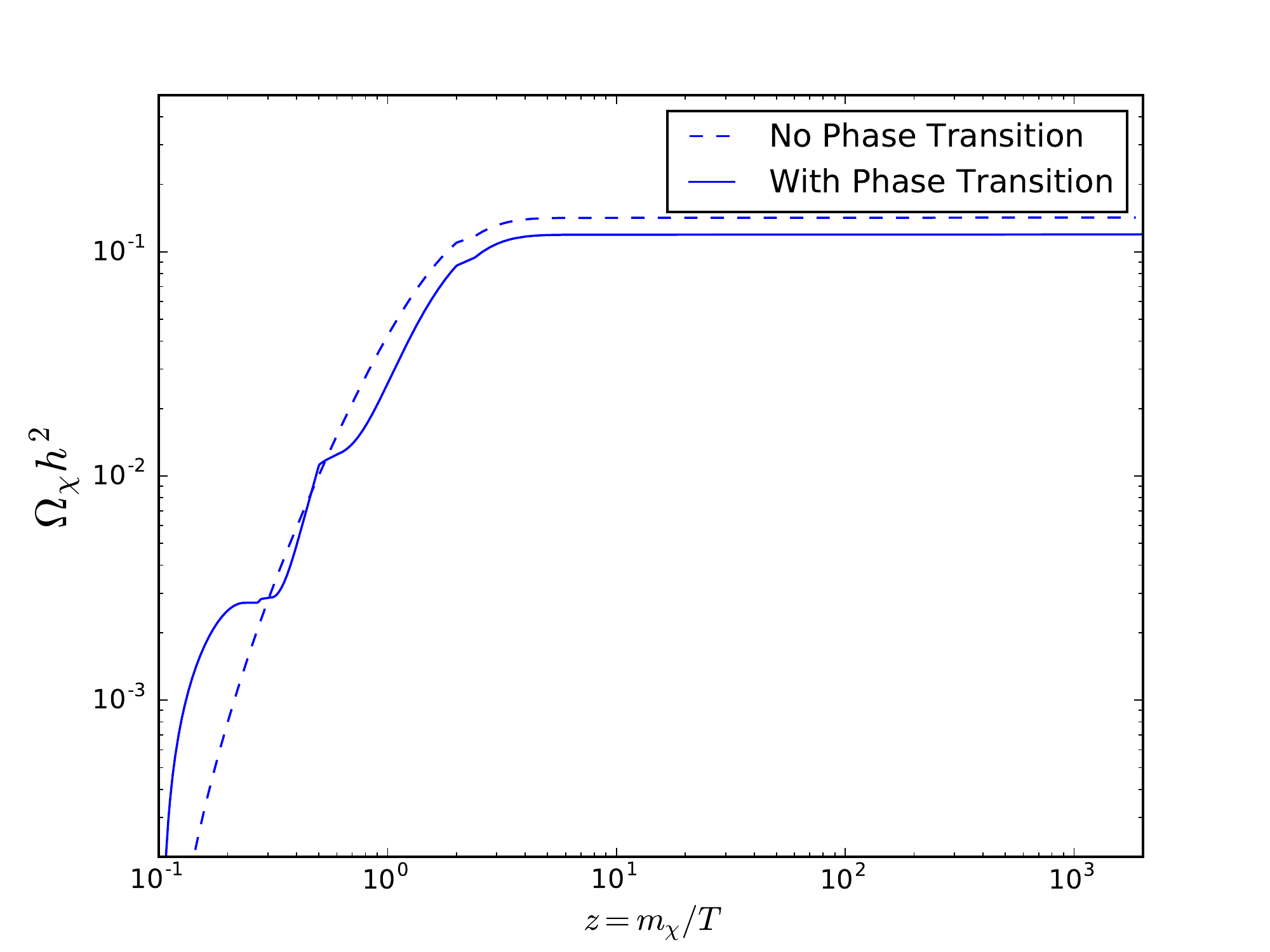}
\includegraphics[width=0.3\textwidth]{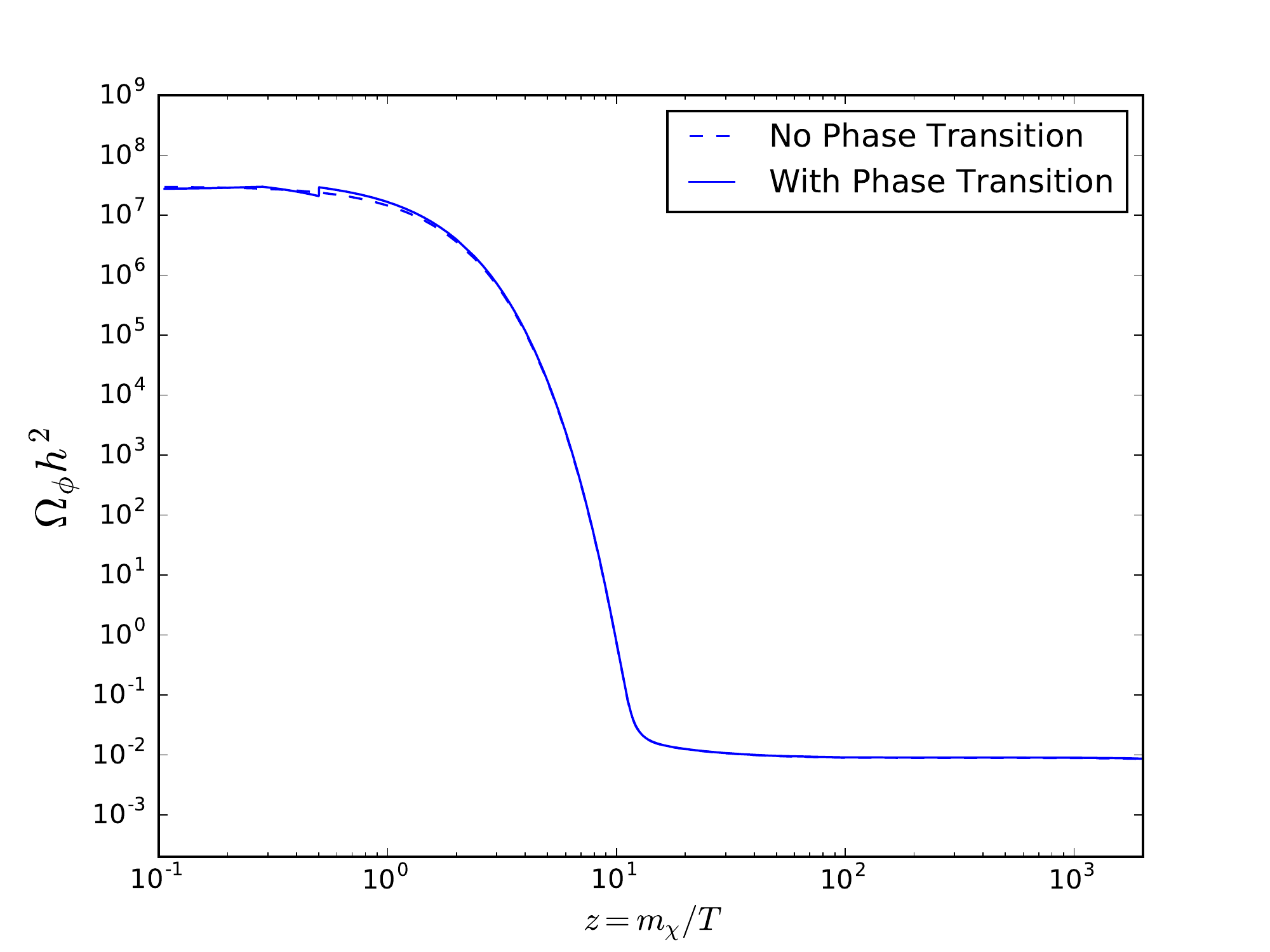}
\includegraphics[width=0.3\textwidth]{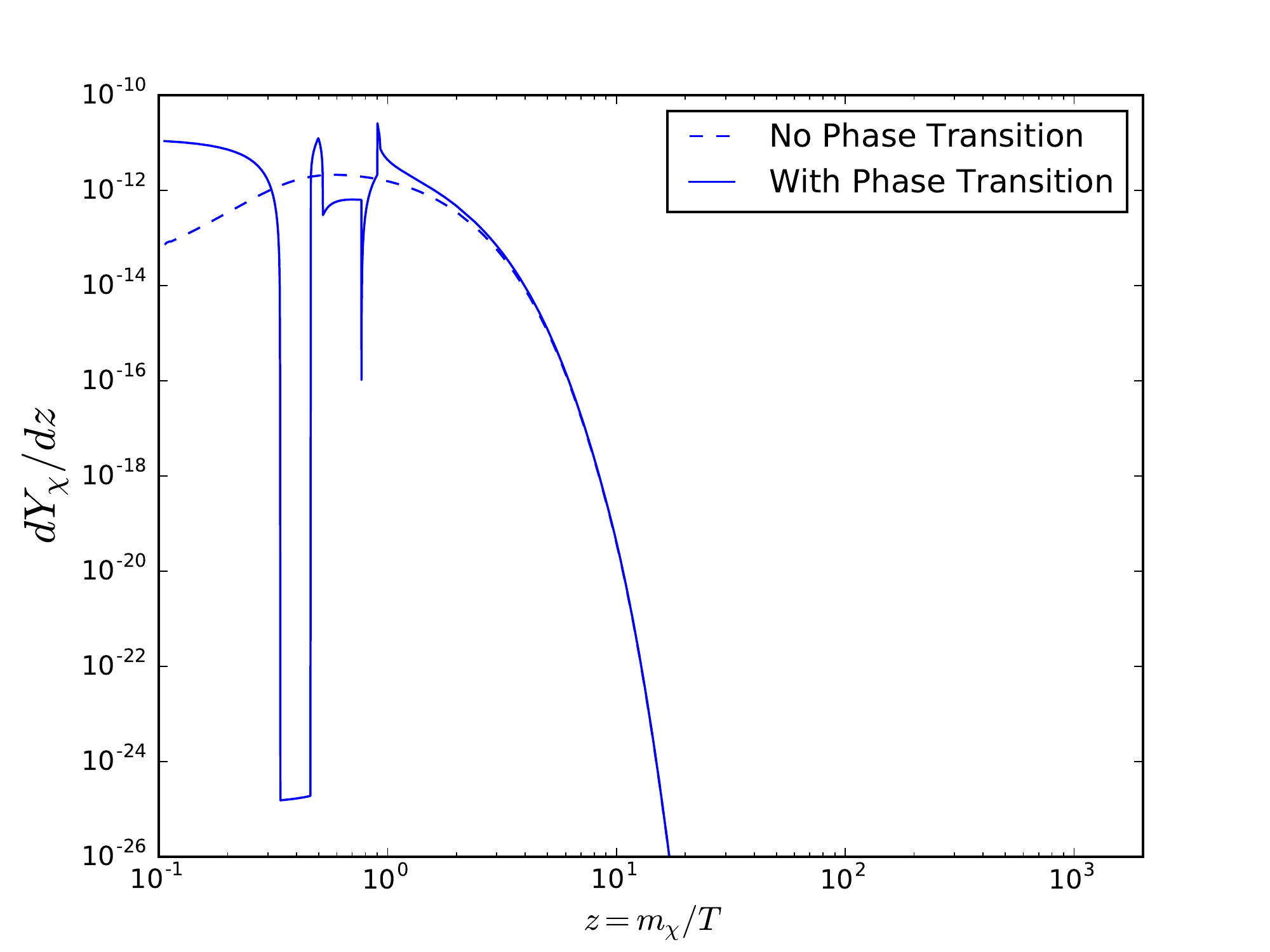}
\includegraphics[width=0.3\textwidth]{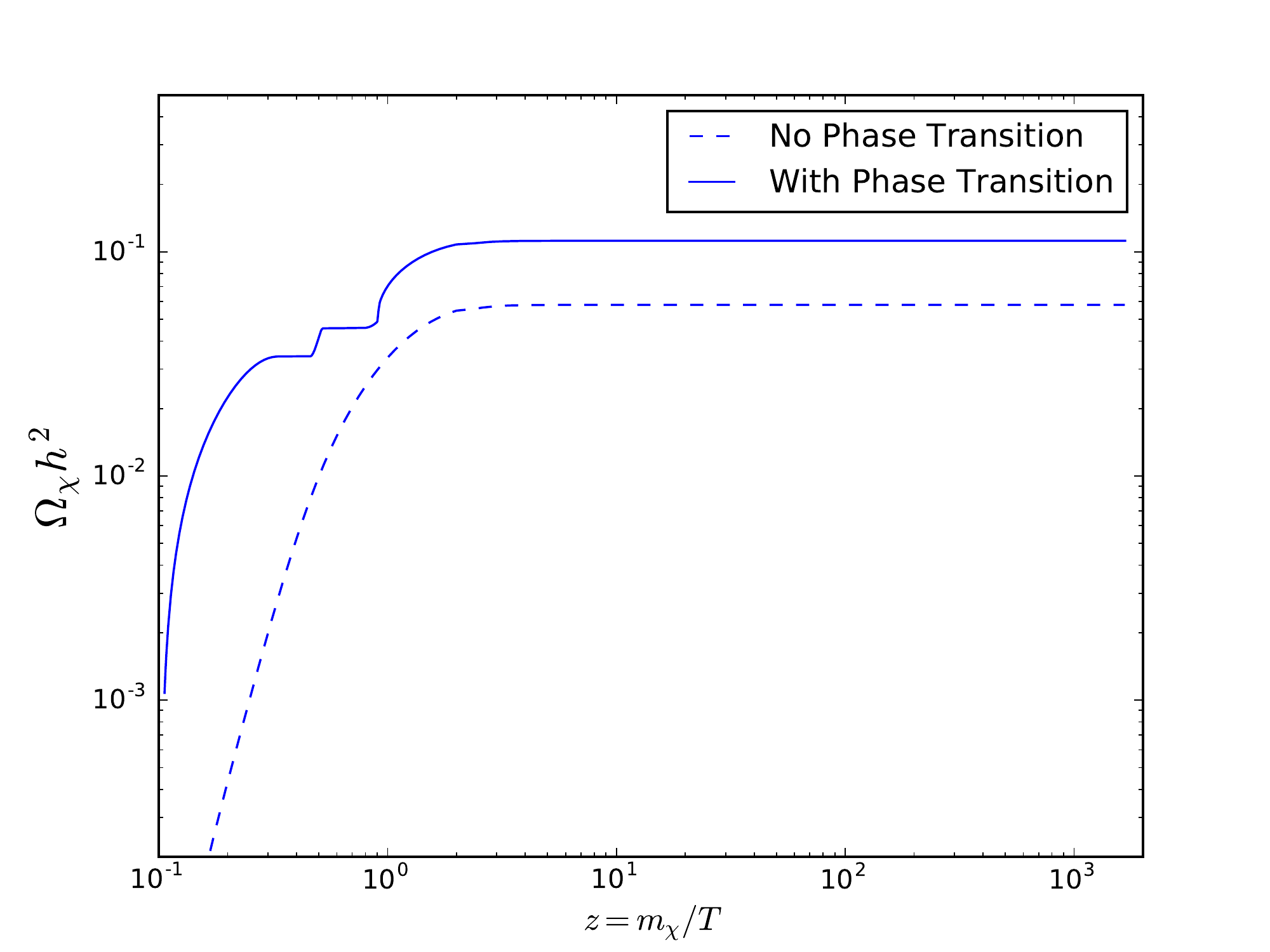}
\includegraphics[width=0.3\textwidth]{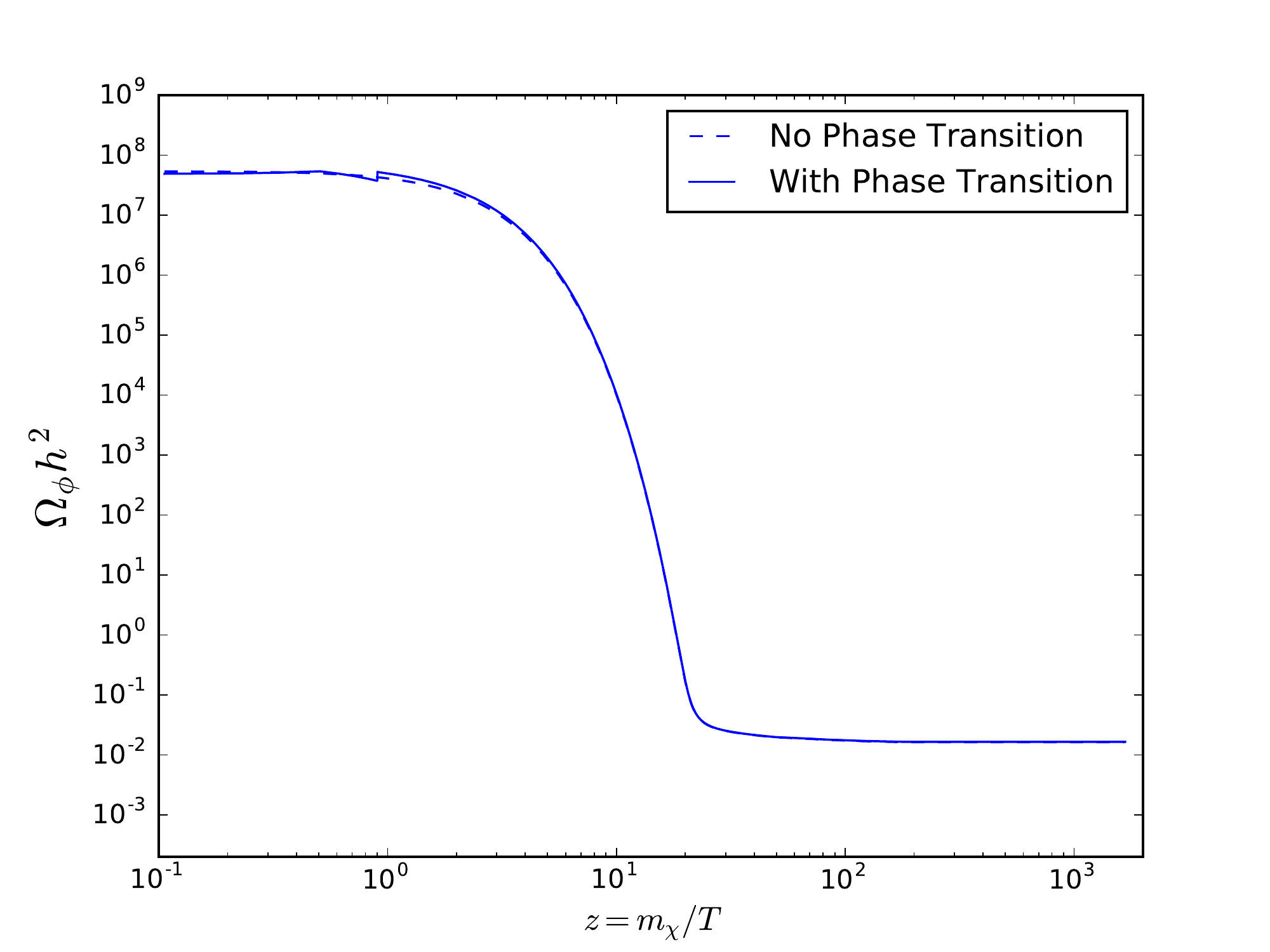}
\caption{The thermal corrected evolution of the number density of $\chi$ and $\phi$ for the case of $m_\phi>m_\chi+m_N$ ( $m_\phi=110$ GeV, $m_\chi=50$ GeV, and $m_N=20$ GeV, $y_D=2.8\times 10^{-12}$), and $m_\phi<m_\chi+m_N$ ($m_\phi=110$ GeV, $m_\chi=90$ GeV, and $m_N=40$ GeV, $y_D=7.3\times 10^{-12}$).}
\label{fig:FIMPchi}
\end{center}
\end{figure}

\begin{itemize}
\item When $z \ll 1$, as the temperature is extremely high, $v_{\phi}(T)=0$ and $\phi$ acquires a significant thermal mass through the Eqn.~(\ref{CCPhi}). This induces $m_{\phi}(T) > m_{\chi}+m_{N}$, and the main contribution to the freeze-in processes is the $\phi \leftrightarrow \chi+N$ channel.

\item The mass of the $\phi$ decreases as the temperature drops. Once $|m_{\chi}-m_{N}| < m_{\phi}(T) < m_{\chi}+m_{N}$, the $1 \leftrightarrow 2$ processes are closed. Much smaller $2 \leftrightarrow 2$ channels dominates the freeze-in processes. This causes the first dip at the left panels in the Fig.~\ref{fig:FIMPchi}.

\item The mass of the $\phi$ continues to decrease until $m_{\phi}(T) < m_{\chi}-m_{N}$, $1 \leftrightarrow 2$ channels reopen, however become $\chi \leftrightarrow \phi + N$ for this time. Therefore, $d Y_{\chi}/d z$ arises again as can be learned from the Fig.~\ref{fig:FIMPchi}. 

Note that after this period, a second-order phase transition happens and $v_{\phi}(T)$ becomes nonzero. After this phase transition, $m_{\phi}(T)$ will rise again as $v_{\phi}(T)$ arises.

\item As the $m_{\phi}(T)$ arises and reaches the $|m_{\chi}-m_{N}| < m_{\phi} (T)< m_{\chi}+m_{N}$ range again, there becomes a second dip at the  left panels in the Fig.~\ref{fig:FIMPchi}. $2 \leftrightarrow 2$  channels dominate the freeze-in processes again, however due to the absent of the $Z_2$ symmetry on this stage, the mixing between the $\phi$ and the SM-Higgs boson introduces much larger SM coupling constants in the $2 \leftrightarrow 2$ diagrams. Therefore, this dip becomes much shallower than the first one.

\item When the $m_{\phi}(T)$ arises above the $m_{\chi}+m_{N}$ again, the $\phi \leftrightarrow \chi + N$ opens again and the $d Y_{\chi}/d z$ recovers. However, in the Fig.~\ref{fig:FIMPchi}, there is a third fake dip before the final recovery. This is because we have dropped all of the $2 \leftrightarrow 2$ channels with an on-shell t/u above the $\phi \leftrightarrow \chi + N$, which is not a very good approximation around the thresholds. Fortunately, this happens within an extremely small period of time, so the final relic abundance results will not be seriously disturbed.

When the temperature drops to $T_n$,  a first-order phase transition happens.

\item A final discontinuity/dip at the left panels of the Fig.~\ref{fig:FIMPchi} appears due to the first-order phase transition, where one have  $|m_{\chi}-m_{N}| < m_{\phi}(T) < m_{\chi}+m_{N}$ and the $1\leftrightarrow2$ process are closed, and again a sub-dominate process  $\phi \chi \leftrightarrow N h$ take a role here. After the $m_\phi$ raise above $ m_{\chi}+m_{N}$ a tiny increase of $dY_\chi/dz$ shows up due to the $1\rightarrow 2$ process is active. 

\item As can be imagined, with the decrease of the number density of the $\phi$, as shown in the top-right panel of Fig.~\ref{fig:FIMPchi}, one have a smoothly drop of $dY_\chi/dz$. After $\phi$ freeze out, one have a decay of  $\phi\to \chi N$, which leads to the tiny smooth uplift of $dY_\chi/dz$. 

\end{itemize}

In the top-middle panel of Fig.~\ref{fig:FIMPchi}, we plot the corresponding thermal abundance evolution of the DM $\chi$, where one can find the thermal effects modified case denoted as ``With Phase Transition" reveals the phase transition history of the model, including the second order phase transition of $\phi$ and the strongly first order phase transition of $h$.  The results of ``With Phase Transition"  and ``No Phase Transition" are close to each other, which is because that the phase transitions of second order and first order occurs around $z\sim m_\chi/T_n\sim 0.5$ and the final and the largest increase of the $\Omega_\chi h^2$ occurs after the phase transitions.

We show the evolution history in the bottom panel of Fig.~\ref{fig:FIMPchi} for the case of $m_{\chi} + m_{N} > m_{\phi}$, and $m_{\chi} < m_{\phi}$. In this scenario, the $\phi \leftrightarrow \chi + N$ process is kinematically prohibited in the h-vacuum shortly after the first-order phase transition. Although $\phi \rightarrow \chi + \nu$, where $\nu$ indicates a active neutrino, becomes the dominant channel, this is suppressed by the squared mixing angle of light and sterile neutrinos $\theta^2$ (see Eq.~\ref{eq:mixang}), and we do not include its negligible contributions in our calculations.  All the thermal steps are basically similar to the top panel, except that shortly after the last sharp leap in the left panel, where the first-order phase-transition happens, the $d Y_{\chi}/d z$ rapidly drops after $m_{\phi}(T)$ becomes larger than $m_{\chi} - m_{N}$. Then, $d Y_{\chi}/d z$ is dominated by the $\phi \chi \leftrightarrow N h$ process and smoothly drops down. Notice that since the $\phi \leftrightarrow \chi + N$ processes are absent shortly after the first-order phase-transition (as can be found in Fig.~\ref{fig:FIMPmphi} when one have $m_\chi+m_N=130$ GeV (the parameter set of the bottom panel of Fig.~\ref{fig:FIMPchi}) instead of 70 GeV (the parameter set of the top panel of Fig.~\ref{fig:FIMPchi})), we need a larger coupling constant $y_D$ compared with the Fig.~\ref{fig:FIMPchi} for a correct dark matter relic abundance. Therefore, one can expect a larger discrepancy between the DM relic density calculated with and without the thermal effects. We stress that, different from the first benchmark shown in the top panel, 
\begin{itemize}
\item The first step second order phase transition of $\phi$ and the first order phase transition of $h$ all happens around the $z\sim m_\chi/T_n\sim 1$. 

\item A large increase of $\Omega_\chi h^2$ happens before the second order phase transition of $\phi$ in comparison with the the top panels case, mostly due to a larger $y_D$ and $z\sim m_\chi/T_n\sim 1$.

\item A significant accumulation of the thermal abundance of $\Omega_\chi h^2$ happens before the decrease of the number density of $\phi$, which can be found through comparison of the bottom-middle and bottom-right panel plots.

\item  Another difference of the two benchmark scenario can be found through the comparison of the top-left panel and the bottom-left panel, no uplift of the $dY_\chi/dz$ shows up in the bottom panel's benchmark due to the $\phi \leftrightarrow \chi + N$ process is not active after the $\phi$ freeze-out for the case of ``With Phase Transition" (indeed the $\phi \leftrightarrow \chi + N$ process only active before the second-stage first-order phase transition) and never active for the case of ``No Phase Transition".

\end{itemize}

\begin{figure}[!ht]
\begin{center}
\includegraphics[width=0.3\textwidth]{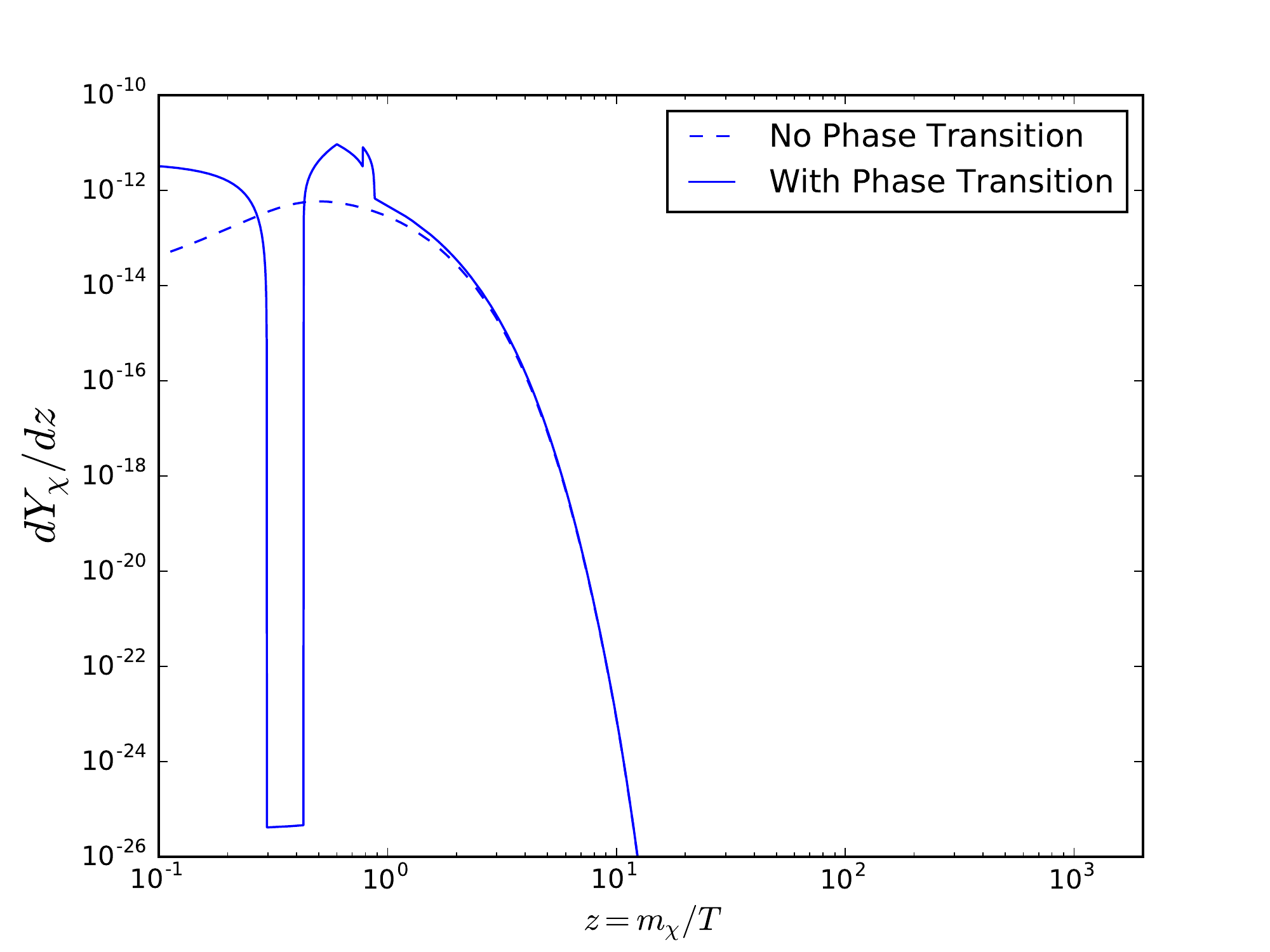}
\includegraphics[width=0.3\textwidth]{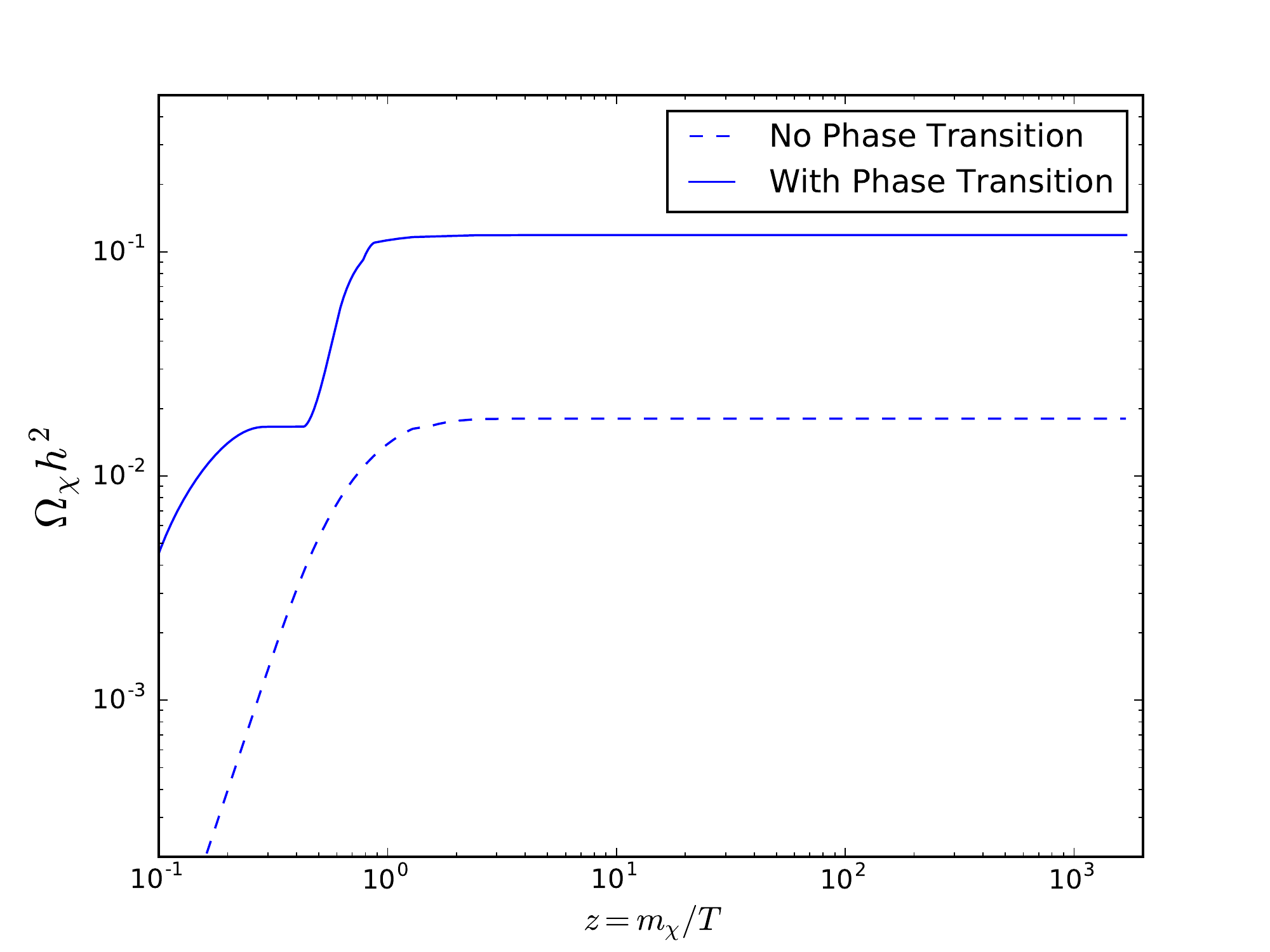}
\includegraphics[width=0.3\textwidth]{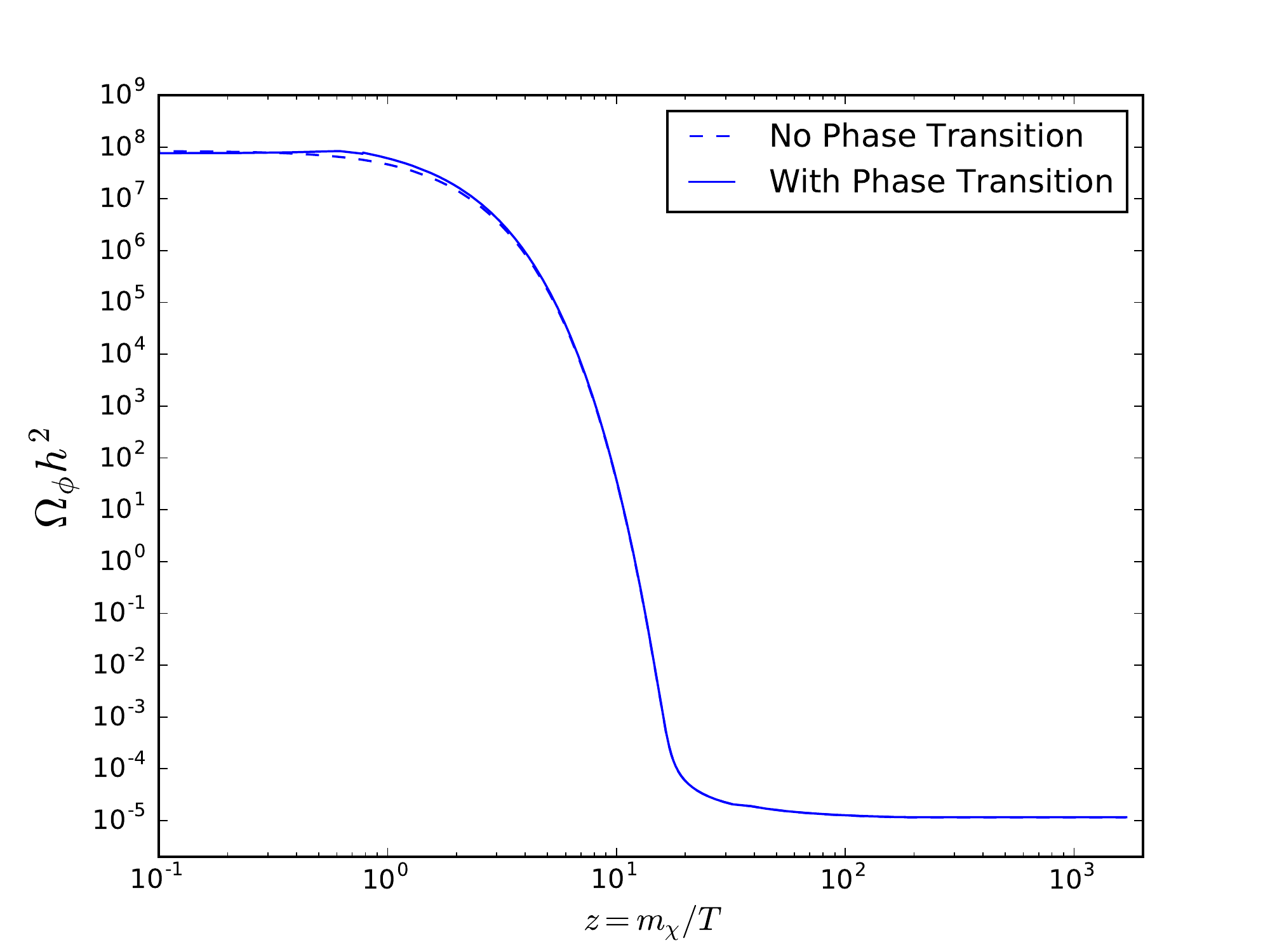}
\caption{The thermal corrected evolution of the number density of $\chi$ and $\phi$ for BP3 ($m_\phi=170$ GeV, $m_\chi=140$ GeV, and $m_N=40$ GeV, $\lambda_{h\phi}=0.65$, $y_D=4.0\times 10^{-12}$.}
\label{fig:FIMPchiBP3}
\end{center}
\end{figure}

We note that
for the $m_N+m_{\chi} > m_{\phi}$ case,  the suppressed $\chi+\nu$ decay width will be very likely to exceed the BBN starting time scale (~1s). In this case, the following decay of the freeze-out $\phi$ can potentially disturb the BBN and even CMB by its decay and injecting particles into the plasma during these epochs. However, if $m_{\phi} > m_h$, and the freeze-out abundance $\Omega_{\phi} h^2$ can be significantly reduced to be $\lesssim 10^{-5}$ (as shown in Fig.~\ref{fig:FIMPchiBP3}), such an influence can be highly reduced by the small $\phi$ abundance.  
In the Fig.~\ref{fig:FIMPchiBP3}, the $d Y_{\chi}/d \chi$ had completely disappeared shortly after z$>$10. This is because we have omitted the $\phi \rightarrow \chi + \nu$ channels in our calculations, which depend on $\theta^2$ and does not affect the final results of $\Omega_\chi h^2$. The same reason that result in the ``With Phase Transition"  behavior of $\Omega_\chi$ (see bottom panels of Fig.~\ref{fig:FIMPchi}) leads to the large discrepancy between the DM relic densities ($\Omega_\chi h^2$) of ``With Phase Transition" and ``No Phase Transition" scenarios (see the middle plot of Fig.~\ref{fig:FIMPchiBP3}).

\section{Comments on the SFOEWPT and the FIMP DM}

A hint from the previous benchmarks is that there can be one tight connection between the phase transition and the produce of the FIMP DM. In this section, we demonstrate the connection and the possibility to search the feature with GW signals and at colliders.

\subsection{Gravitational wave signals}
\label{sec:GWTn}

One of the crucial parameter for the Gravitational wave is the strength of the phase transition, 
the parameter $\alpha$.
Which is the energy budget of SFOEWPT normalized by the radiative energy, being defined by,
\begin{eqnarray}
\alpha=\frac{\Delta\rho}{\rho_R}\;,
\end{eqnarray}
where the radiation energy of the bath or the plasma background $\rho_R$ is given by
\begin{eqnarray}
\rho_R=\frac{\pi^2 g_\star T_\star^4}{30}\;,
\end{eqnarray}
The $\Delta\rho$ is
the released latent heat ( vacuum energy density or energy budget of SFOEWPT) from the phase transition to the energy density of the radiation bath 
or the plasma background. This is given by the difference of the energy density between the false (here it is $\phi$ vacuum, $\rho(\phi_n,T)$) and the true vacuum (the $h$-vacuum or EW broken vacuum, $\rho(v_n,T)$), 
\begin{eqnarray}
\rho(\phi_n,T_n)&=& -V(\phi,T)|_{T=T_n}+T\frac{d\, V(\phi,T)}{d\,T}|_{T=T_n}\;,
\\
\rho(v_n,T_n)&=& -V(h,T)|_{T=T_n}+T\frac{d\, V(h,T)}{d\,T}|_{T=T_n}\;.
\end{eqnarray}
Another crucial parameter of $\beta$ characterizes the inverse time duration of the SFOEWPT and thus the GW spectrum peak frequency,
\begin{eqnarray}
\frac{\beta}{H_n}=T\frac{d (S_3(T)/T)}{d T}|_{T=T_n}\; ,
\end{eqnarray}
with $H_n$ being the Hubble constant at the bubble nucleation temperature $T_n$.

The gravitational wave signals generated by the SFOEWPT mainly include three 
sources: bubble collisions, sound waves and Magnetohydrodynamic 
turbulence (MHD)  in the plasma~\cite{Caprini:2015zlo,Cai:2017cbj}. The total energy spectrum of 
the three sources is given by,
\begin{eqnarray}
  \Omega_{\text{GW}}h^2 \simeq \Omega_{\text{col}} h^2 + \Omega_{\text{sw}} h^2 + \Omega_{\text{turb}} h^2\;.
\end{eqnarray}

The first source of the gravitational waves from the bubble collision estimated using the envelop approximation~\cite{Kosowsky:1991ua,Kosowsky:1992rz,Kosowsky:1992vn} is~\cite{Huber:2008hg},
\begin{eqnarray}
  \Omega_{\text{col}} h^2 = 1.67\times 10^{-5} \left(\frac{H_{\ast}}{\beta}\right)^2
  \left(\frac{\kappa \alpha}{1+\alpha}\right)^2 \left( \frac{100}{g_{\ast}} \right)^{1/3}
  \left( \frac{0.11 v_w^3}{0.42 + v_w^2} \right) 
  \frac{3.8(f/f_{\text{env}})^{2.8}}{1+2.8(f/f_{\text{env}})^{3.8}} \ , 
\end{eqnarray}
where the bubble wall velocity $v_w$ and the efficient factor $\kappa$ that characterizing the fraction of latent heat
deposited in a thin shell, are all functions of the parameter of $\alpha$~\cite{Kamionkowski:1993fg},
\begin{eqnarray}
v_w \simeq \frac{1/\sqrt{3} + \sqrt{\alpha^2+2\alpha/3} }{1+\alpha}\;,  \nonumber \quad \quad
\kappa \simeq  \frac{0.715\alpha + \frac{4}{27} \sqrt{3\alpha/2}}{1+0.715\alpha} \;,
\end{eqnarray}
and the peak frequency $f_{\text{env}}$ is,
\begin{eqnarray}
  f_{\text{env}} = 16.5\times 10^{-6} \left(\frac{f_{\ast}}{H_{\ast}}\right)  \left(\frac{T_{\ast}}{100\text{GeV}}\right)
  \left(\frac{g_{\ast}}{100}\right)^{1/6}
  \text{Hz}\;.
\end{eqnarray}
The second and the third important sources of the sound waves and the MHD are,
\begin{eqnarray}
  \Omega_{\text{sw}} h^2 & =& 2.65\times 10^{-6} \left(\frac{H_{\ast}}{\beta}\right)
  \left(\frac{\kappa_v \alpha}{1+\alpha}\right)^{2} \left(\frac{100}{g_{\ast}}\right)^{1/3} v_w 
  \left(\frac{f}{f_{\text{sw}}}\right)^3 \left( \frac{7}{4+3(f/f_{\text{sw}})^2} \right)^{7/2} \;,\nonumber\\
  \Omega_{\text{turb}} h^2  &= &3.35\times 10^{-4} \left(\frac{H_{\ast}}{\beta}\right)
  \left(\frac{\kappa_{\text{turb}} \alpha}{1+\alpha}\right)^{3/2} 
  \left(\frac{100}{g_{\ast}}\right)^{1/3} v_w 
  \frac{(f/f_{\text{turb}})^3}{[1+(f/f_{\text{turb}})]^{11/3} (1+8\pi f/h_{\ast})} \;.\nonumber\\
\end{eqnarray}
Here, the fraction of latent heat transformed into the bulk motion of the fluid for sound waves and MHD are given by $\kappa_v \approx \alpha(0.73+0.083\sqrt{\alpha} + \alpha)^{-1}$  and $\kappa_{\text{turb}} \approx 0.1 \kappa_v$; the peak frequency of sound waves and MHD are, 
\begin{eqnarray}
 && f_{\text{sw}} = 1.9\times 10^{-5} \frac{1}{v_w} 
  \left(\frac{\beta}{H_{\ast}}\right) \left(\frac{T_{\ast}}{100\text{GeV}}\right) 
  \left(\frac{g_{\ast}}{100}\right)^{1/6} \text{Hz} \;,\\
 &&   f_{\text{turb}} = 2.7\times 10^{-5} \frac{1}{v_w} 
  \left(\frac{\beta}{H_{\ast}}\right) \left(\frac{T_{\ast}}{100\text{GeV}}\right) 
  \left(\frac{g_{\ast}}{100}\right)^{1/6} \text{Hz} \;;
\end{eqnarray}
the Hubble parameter at present is given by 
\begin{eqnarray}
h_{\ast}=1.65\times 10^{-2} {\text{mHz}}\frac{T_{\ast}}{100 \text{GeV}}(\frac{g_{\ast}}{100})^{1/6}\;.
\end{eqnarray}

\begin{figure}[!ht]
\begin{center}
\includegraphics[width=0.4\textwidth]{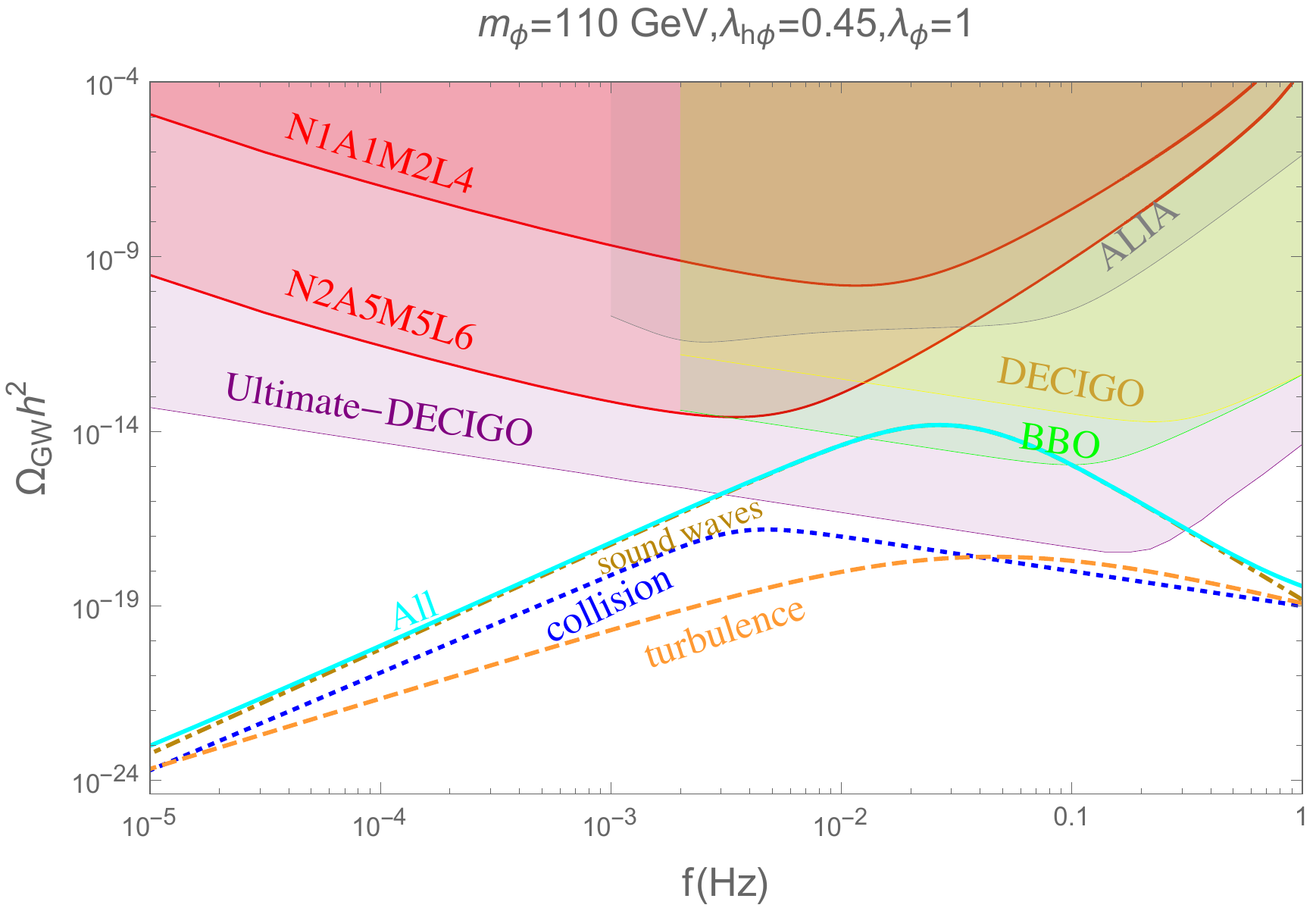}
\includegraphics[width=0.4\textwidth]{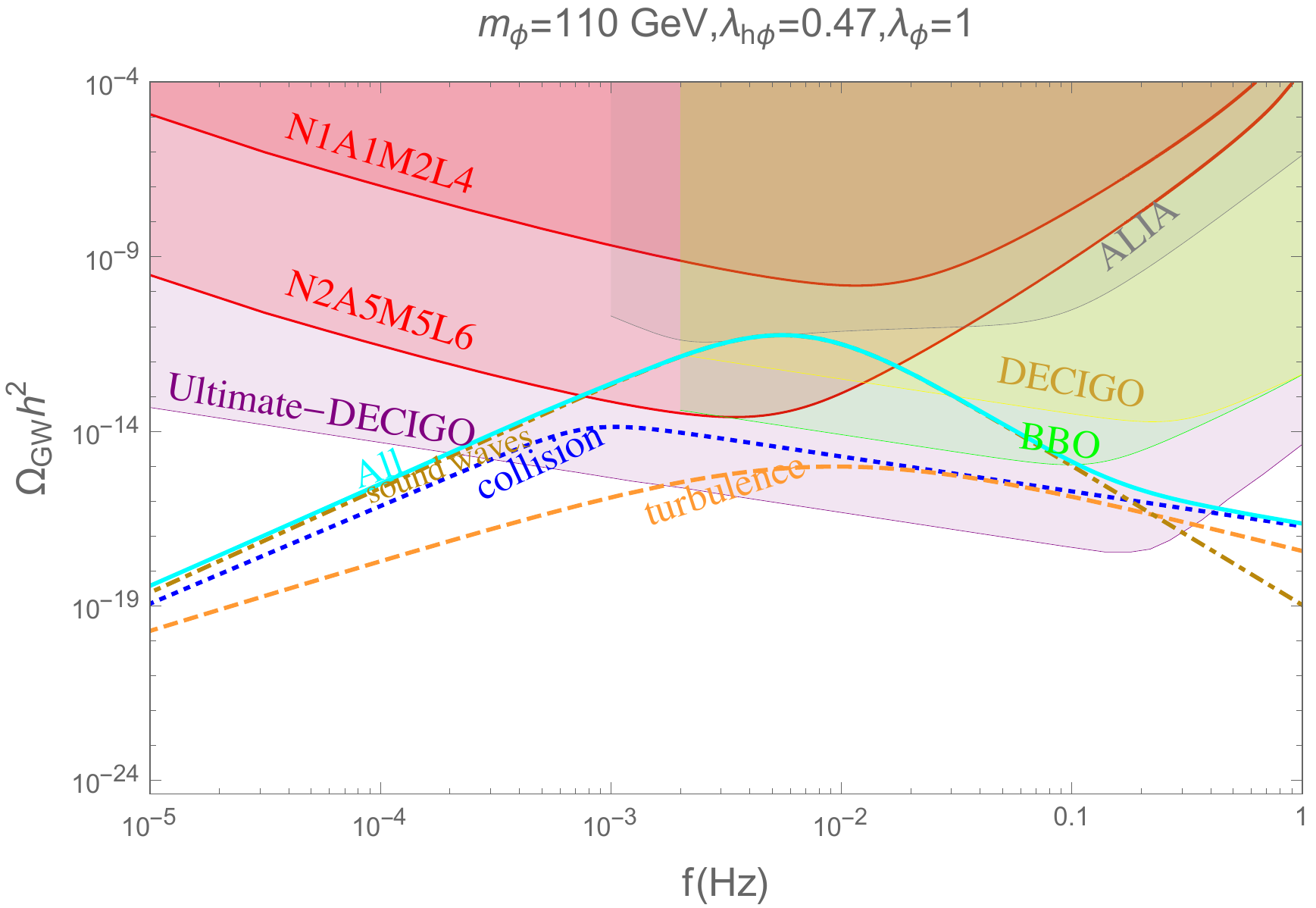}
\includegraphics[width=0.4\textwidth]{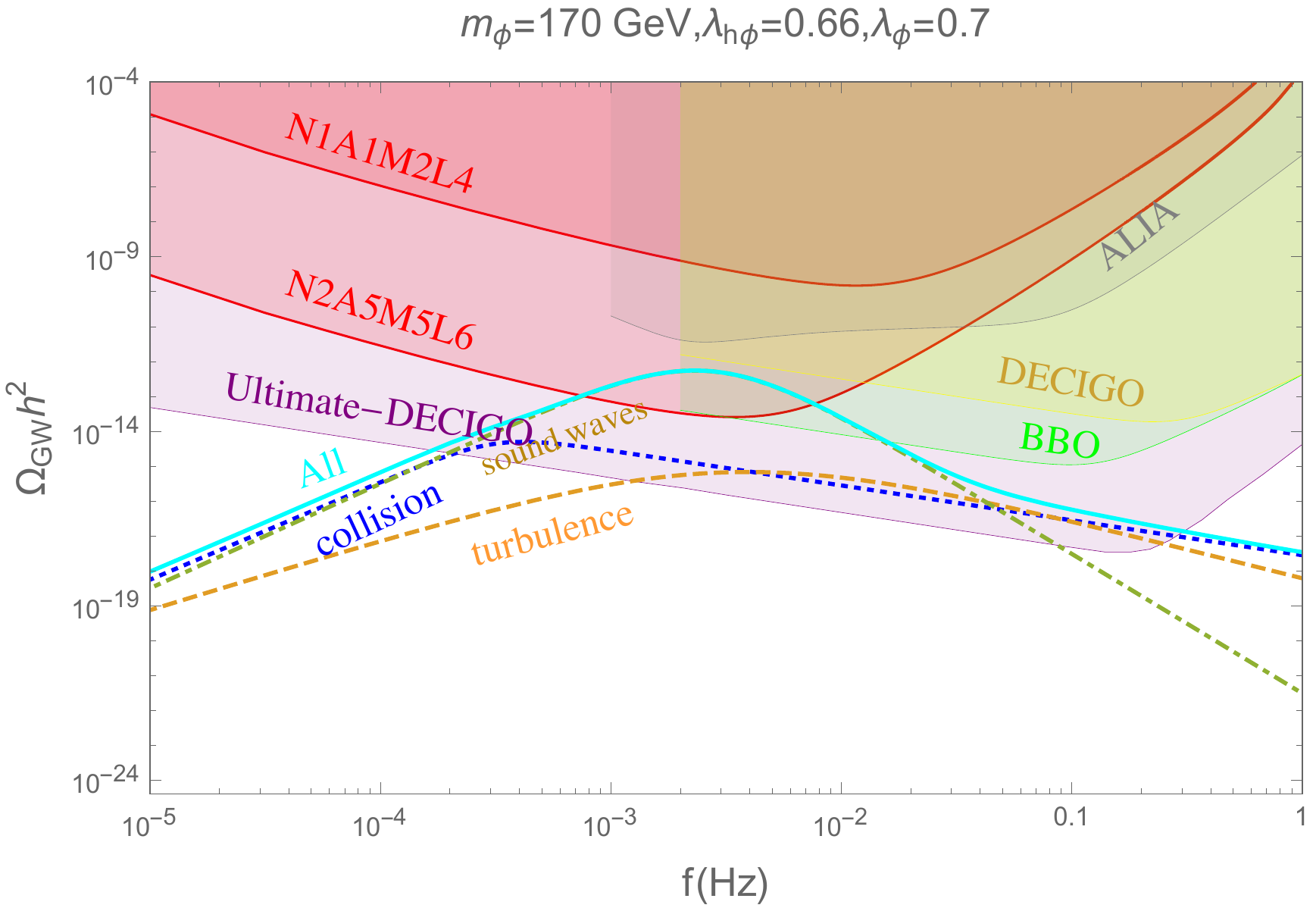}
\includegraphics[width=0.4\textwidth]{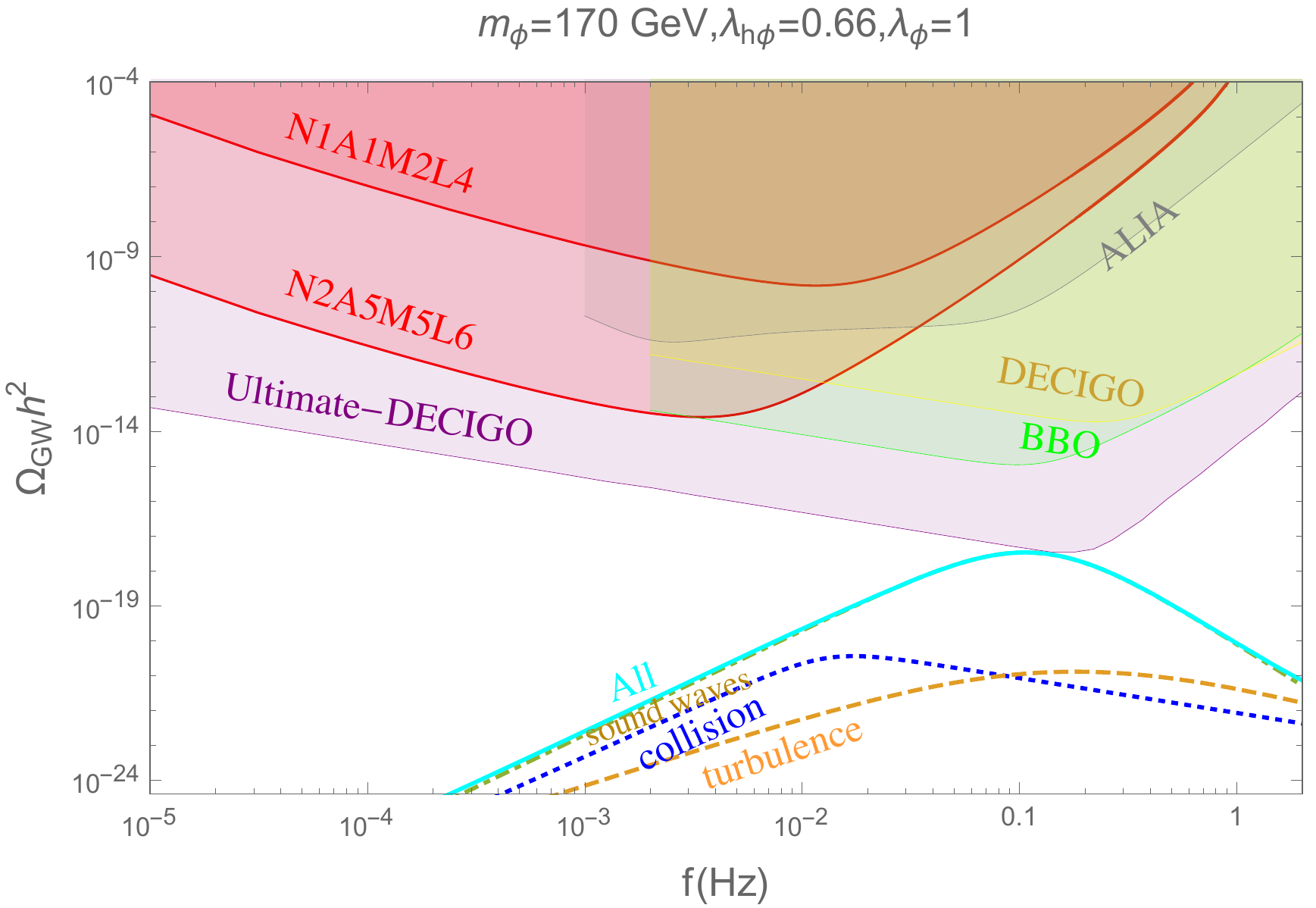}

\caption{The GWs generated by the SFOEWPT from three main sources:
sound waves(brown dotdashed line), collision(blue dotted line), turbulence(orange dashed line) and total
contribution(cyan solid line). The colored regions depict the experimental sensitivities of
eLISA(two configurations with notation NiAjMkLl), ALIA(gray), BBO(green), DECIGO(yellow) and
Ultimate-DECIGO(purple)}
\label{fig:GW}
\end{center}
\end{figure}

In Fig.~\ref{fig:GW}, we show the possibility to probe the benchmarks of the Fig.~\ref{fig:FIMPchi} and Fig.~\ref{fig:FIMPchiBP3} with the GW detectors. 
The sensitivities for the proposed space based interferometers: LISA~\cite{Audley:2017drz} with two design 
configurations in notation NiAjMkLl~\cite{Caprini:2015zlo,Klein:2015hvg}, BBO, 
DECIGO (Ultimate-DECIGO)~\cite{Kudoh:2005as} and ALIA~\cite{Gong:2014mca} are shown with different color shaded regions. 
 For the GW signals produced by the SFOEWPT around the benchmark of the Fig.~\ref{fig:FIMPchi} is show in the top panels of Fig.~\ref{fig:GW}.
A larger scalar quartic coupling $\lambda_{h\phi}$ can leads to a larger $\Delta\rho$, and therefore the 
corresponding GW signal from the SFOEWPT is of high magnitude and can be probed easier. We note that a tiny increase of $\lambda_{h\phi}$ drives a lower peak frequency.
In the lower panel of the Fig.~\ref{fig:GW}, we plot the GW signals from the SFOEWPT around the benchmarks of Fig.~\ref{fig:FIMPchiBP3}, the feature is that a smaller $\lambda_\phi$ leads to a lower peak frequency and higher magnitude of the spectrum, which is because that a smaller $\lambda_\phi$ can induce a larger $\Delta\rho$ and a lower $\beta/H$.
The study of dynamics of phase transition and the produced GW signals with the same scalar sectors under $Z_2$ symmetry can be found in Ref.~\cite{Beniwal:2017eik,Vaskonen:2016yiu,Kurup:2017dzf}.

\subsection{On the GW signals and FIMP DM production}

From the side of DM, one ingredient is the open/close of the $1\leftrightarrow 2$ process at zero temperature, which is determined by the mass threshold at the zero temperature. If it is not kinematic allowed at zero temperature, the thermal effects would make it active before or during the phase transition process, and therefore results in a significant difference thermal relic abundance in comparison with the traditional one without taking into account the thermal effects. Another ingredient is that when there is a larger portal coupling,     
there can be a larger contribution to DM production from the annihilation cross section of $2\leftrightarrow2$ process $\chi\phi\leftrightarrow Nh$. 

Furthermore, a larger $m_\phi$( or $\lambda_{h\phi}$) leads to the second order phase transition to occur much latter (or earlier) that can decrease (or increase) the contribution of decay/inverse decay process to the FIMP DM productions, especially when the decay/inverse decay channel is not open at zero temperature as the benchmark scenario of the bottom panel in Fig.~\ref{fig:FIMPchi}. This is due to the interval between the reheating temperature $T_R$ and $T_\phi$ is decreased (increased), and therefore a decrease (increase) of the DM abundance's accumulation. 

\subsection{Collider interaply} 

In the scenarios being explored in this work, we do not expect to search for the $\chi$ field at colliders
due to its extremely long decay length far beyond the scope of the detectors.
The other two new particles beyond the SM are the dark scalar field $\phi$ and the sterile neutrino that are relevant for the neutrino mass generation.

\begin{figure}[!ht]
\begin{center}
\includegraphics[width=0.6\textwidth]{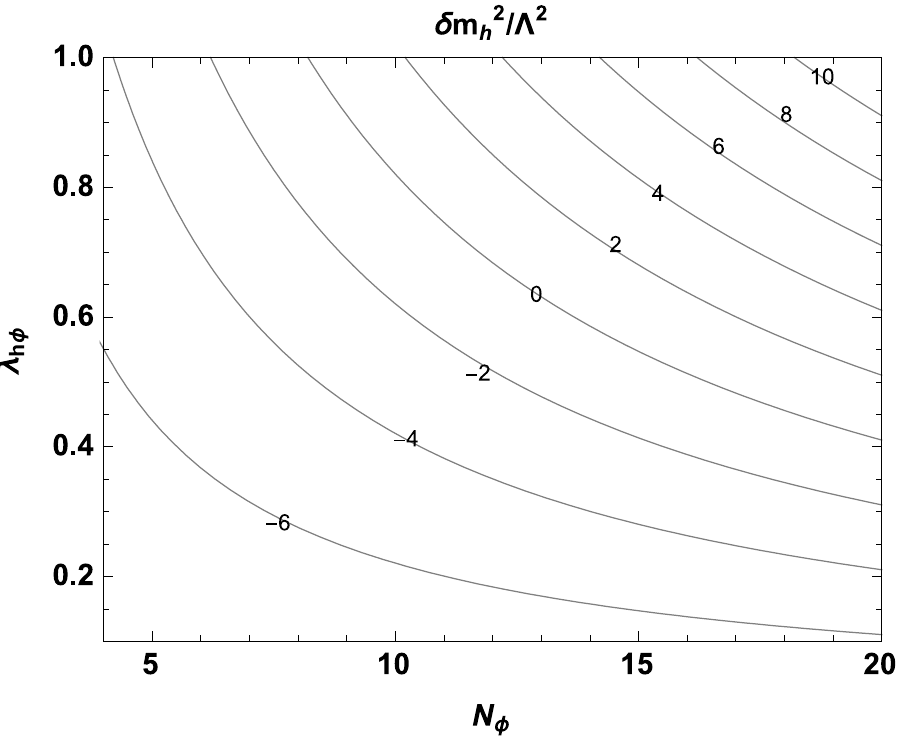}
\caption{The quadratic corrections in the parameter space of $N_\phi$ and $\lambda_{h\phi}$.}
\label{fig:naturalness}
\end{center}
\end{figure}

With the gauge invariant approach by taking into account the tadpole contributions to the Higgs two-point green's function~\cite{Bian:2013xra,Bian:2013wna,Bian:2014cja}, the $\phi$ introduced here can be related with the fine-tuning of the Higgs, the quadratic corrections from the hidden scalar singlet is given by,
\begin{eqnarray}
\delta m_h^2=\frac{1}{16\pi^2}(12\lambda-4 N_c y_t^2+\frac{9}{2}g_2^2+\frac{3}{2}g_1^2+\lambda_{h\phi})\Lambda^2\;.
\end{eqnarray}
Supposing there are N scalars of $\phi$, then we have,
\begin{eqnarray}
\delta m_h^2=\frac{1}{16\pi^2}(12\lambda-4 N_c y_t^2+\frac{9}{2}g_2^2+\frac{3}{2}g_1^2+N\lambda_{h\phi})\Lambda^2\;.
\end{eqnarray}
We show the value of $\delta m_h^2/\Lambda^2$ in the panel of $N_\phi$ and $\lambda_{h\phi}$, see Fig.~\ref{fig:naturalness}.
In the situation, we can expect the FIMP DM production to be more efficient, and to produce the correct DM relic abundance one just need a $y^{N_\phi}\sim y_\chi/\sqrt{N}$. More exactly, $y^{N_\phi}_\chi\sim y_\chi/\sqrt{18}$ in the parameter region where one can obtain the cancellation of $\delta m_h^2$ to alleviate the Hierarchy problem where the GW signals can be probed (in this work we use the benchmarks with $\lambda_{h\phi}\sim 0.5$) with the phase transition not to be affected by the number of scalars for the two-step pattern~\cite{Cheng:2018axr}. 
The GWs signatures being explored in this work mostly focus on the case of $m_\phi>m_h/2$, and therefore, we expect the benchmarks can be probed by the off-shell Z-pair search at LHC~\cite{Goncalves:2017iub}. 
For the search of this scenario at future linear colliders including the CEPC, ILC, and Fcc-ee we refer to Ref.~\cite{Curtin:2014jma,Cheng:2018axr}. The study of Ref.~\cite{Curtin:2014jma} illustrate that the Fcc-ee and 100 TeV pp collider are complementary and of great
potential to probe this kind of phase transition, the numbers of $N_\phi$ might be able to be determined. 

For the typical mass of the RHN being adopted in this work, we have the decay length being given by~\cite{Izaguirre:2015pga},
\begin{eqnarray}
&&\Gamma(N\to \ell_\alpha^- \ell_\beta^+ \nu_\beta)=\frac{G_F^2 m_N^5 \theta_{ij}^2}{192\pi^3}\;,\\
&&\Gamma(N\to \ell_\alpha^- \ell_\alpha^+ \nu_\alpha)=\frac{G_F^2 m_N^5 \theta_{ij}^2}{768\pi^3}(1+4\sin^2\theta_W+8\sin^4\theta_W)\;,
\end{eqnarray}
For the typical yukawa $y_N$ and mixing angle between the sterile neutrino and active neutrino ($\theta_{ij}$), with the mixing angle being around $\theta_{ij}^2\sim \mathcal{O}(10^{-11}-10^{-9})$ the decay length is estimated to be $c\tau_N\sim \mathcal{O}(10)$ m, which is beyond the scope of LHC trilepton and lepton jet search performed in Ref.~\cite{Izaguirre:2015pga}. The proposed SHiP experiment is also unable to probe the scenario due to its proposed mass range~\cite{Deppisch:2015qwa}.

\section{Conclusions}

In this paper, we implement the thermal history of symmetry change with the Universe cools down (phase transition) into the dark matter production history.
With the assistance of right handed sterile neutrino that lives in thermal bath when the FIMP DM is producing, the dark matter production process can be multi-step due to the effects of the EWPT. More precisesly, the thermal history of the EWPT
can change the kinematic threshold of DM production process because the thermal corrected mass and the VEVs of $h$ and $\phi$ are all depends on the temperature. We studied the two-step EWPT impacts, all the phase transition history can be revealed by the relic density evolution with the 
temperature drops, including the second order phase transition of the $Z_2$ symmetry and the first order phase transition from the $Z_2$ symmetry broken phase with the the EW symmetry to the EW symmetry broken phase with the $Z_2$ symmetry being respected. Here, both the decay length of the FIMP DM and sterile neutrino are beyond the scope of the present colliders. The interaction rate for the two-step phase transition, i.e., the quartic coupling between the Higgs and the extra scalar $\phi$, can results in the gravitational wave signals to be probed by the future gravitational wave experiments. To alleviate the Hierarchy problem the number of $\phi$ should be around $\sim 18$ with a $y^{N_\phi}_\chi\sim y_\chi/\sqrt{18}$ to yields the correct relic abundance through thermal corrected FIMP mechanism. To address the puzzle of BAU within the EWBG mechanism, an extra CP violation source from the high dimensional operators are necessary, see Ref.~\cite{Espinosa:2011eu,Huang:2018aja,Vaskonen:2016yiu,Jiang:2015cwa,Cline:2012hg}.
 
 \section{acknowledgements}

We are grateful to Lian-Tao Wang, Stefano Profumo and Michael A. Schmidt for communication and discussions on the entropy change during the EWPT process and its effects on the relic density estimation. We thank Oscar Zapata, Probir Roy,
and Takashi Toma for discussions on the non-thermal dark matter production mechanisms; and D. Goncalves, T. Han, and S. Mukhopadhyay for helpful communications on the Higgs naturalness calculations when additional singlets are considered; and Huai-ke Guo for helpful discussions on the GWs from SFOEWPT. 
The work of LGB is Supported by the National Natural Science Foundation of China (under
grant No.11605016 and No.11647307), Basic Science Research Program through the National Research Foundation of Korea (NRF) funded by the Ministry of Education, Science and Technology (NRF-2016R1A2B4008759), and Korea Research Fellowship Program through the National Research Foundation of Korea (NRF) funded by the Ministry of Science and ICT (2017H1D3A1A01014046).
This work of YLT is supported by the Korea Research Fellowship Program through the National Research Foundation of Korea (NRF) funded by the Ministry of Science and ICT (2017H1D3A1A01014127).

\appendix

\section{The neutrino mixing}
\label{sec:nm}
For the practical neutrino mass spectrum and mixing pattern, at least two sterile-neutrinos are required. In the literature, people prefer the three sterile neutrinos. In the pseudo-Dirac sterile neutrino case, a general mass matrix is given by
\begin{eqnarray}\label{eq:seesaw}
M_{\nu, N_R, N_L} = \left[
\begin{array}{ccc}
0 & m_D & \mu_I^{T} \\
m_D^T & \mu_1^{3 \times 3} & m_{N_D}^{3 \times 3} \\
\mu_I & m_{N_D}^{{3 \times 3}  T} & \mu_2^{3 \times 3}
\end{array} \right],
\end{eqnarray}
where all the sub-matrices are $3 \times 3$ mass matrices. $m_{N_D}^{3 \times 3}$, $\mu_{1, 2}^{3 \times 3}$ are the $3 \times 3$ mass matrix extended from the corresponding terms in the Eq.~(\ref{PDLag}). $m_{D i j} = y_{N i j}^{3 \times 3}v/\sqrt{2}$, $\mu_{I i j} = y_{NC i j}^{3 \times 3}v/\sqrt{2}$, where $y_{N(C)}^{3 \times 3}$ are also the  corresponding $3 \times 3$ extensions. Seesaw models require that $\mu_{I, 1, 2}^{3 \times 3} \ll m_D \ll m_{N_D}^{3 \times 3}$, and for a sufficient light neutrino mass, $y_{N i j} \gtrsim 10^{-6}$. In the literature, The $\mu_{1}$ or $\mu_{2} \neq 0$ and $\mu_I=0$ situation  is called the inverse see-saw model. The  $\mu_{1} = \mu_{2} = 0$ and $\mu_I \neq 0$ situation is called the linear see-saw model. Without loss of generality, we diagonalize $m_N$ at first, and $m_{N_D} =\text{diag} \left[ m_{N_D 1}, m_{N_D 2}, m_{N_D 3}  \right]$. Although the light neutrino masses and mixings depend on all of the $m_D^{3 \times 3}$, $m_{N D i}$, $\nu_{I, 1, 2}^{(3 \times 3)}$ elements, the mixings between the light and sterile neutrinos are not sensitive to many of them except
\begin{eqnarray}
(\theta_{i j})^2 = \left( \frac{m_{Dij}^{3 \times 3}}{m_{N_D j}} \right)^2\;,\label{eq:mixang}
\end{eqnarray}
where $i = e$, $\mu$, $\tau$, and $j=1$, $2$, $3$.

The electroweak precision measurements constrain the $\theta_{ij}$ parameters. The most stringent bounds originate from the FCNC processes, and thus constrain the off-diagonal elements in the $m_D$. In the literature, people are interested in the cases when $m_{N, D} \propto I$, therefore the significant FCNC processes are avoided. In this case, all the sterile neutrinos have a unified mass and decay width. 

\section{Two stage phase transition and mixing between $\chi$ and $N$}

For the case that  $\phi$ get a VEV during the phase transition, one have the mixing between the sterile neutrino and the $\chi$. In the Majorana sterile neutrino case,
\begin{eqnarray}\label{eq:seesaw}
\left(\begin{array}{c}
N \\
\chi
\end{array}\right)= \left(
\begin{array}{cc}
\cos\theta & \sin\theta  \\
-\sin\theta &\cos\theta 
\end{array} \right) \left(\begin{array}{c}
\chi_1\\
\chi_2
\end{array}\right).
\end{eqnarray}
with mixing angle being given by 
$\tan(2\theta)=2y_\chi v_\phi(T)/(m_N-m_\chi)$. In the pseudo-Dirac case, things are a little bit complicated. Written in the basis of two-component Weyl spinors, the mass complete mass matrix is given by
\begin{eqnarray}
\left[ \begin{array}{ccc}
\chi^w &  N_1 &  N_2
\end{array} \right] \left[
\begin{array}{ccc}
m_{\chi} & y_{\chi D} v_{\phi}(T) & y_{\chi D} v_{\phi}(T) \\
 y_{\chi D} v_{\phi}(T) & \mu_2 &m_{N_D} \\
 y_{\chi D} v_{\phi}(T) & m_{N_D} & \mu_1
\end{array} \right]
\left[ \begin{array}{c}
\chi^w \\
 N_1 \\
 N_2
\end{array} \right] + \text{h.c.}, \label{Dirac_Mass}
\end{eqnarray}
where $\chi^w$ is the Weyl component of the four-spinor $\chi$. We do not need to fully diagonalize the (\ref{Dirac_Mass}). We only need to rotate out all the terms between the $\chi$ and $N_D$ components. Therefore, we can still treat the rotated sterile neutrino-like component of the fermions as a pair of pseudo-Dirac neutrinos for the simplicity of the calculations. The rotate matrix is perturbatively calculated to be
\begin{eqnarray}
V = \left[ \begin{array}{ccc}
1 & \frac{v_{\phi}(T)}{|m_{\chi} - m_{N_D}|} & \frac{v_{\phi}(T)}{|m_{\chi} - m_{N_D}|} \\
\frac{v_{\phi}(T)}{|m_{\chi} - m_{N_D}|} & 1 & 0 \\
\frac{v_{\phi}(T)}{|m_{\chi} - m_{N_D}|} & 0 & 1
\end{array} \right],
\left[ \begin{array}{c}
\chi^{w \prime} \\
 N_1^{\prime} \\
 N_2^{\prime}
\end{array} \right] 
=
V \left[ \begin{array}{c}
\chi^w \\
 N_1 \\
 N_2
\end{array} \right] 
\end{eqnarray}
up to the first order of $v_{\phi}(T)$. Here $\chi^{w \prime}$, $N_{1,2}^{\prime}$ are the new rotated Dirac spinors, and these can still be combined into a Majorana and a Dirac spinor
\begin{eqnarray}
\chi^{\prime} = \left[ \begin{array}{c}
\chi^{w \prime} \\
i \sigma^2 \chi^{w \prime *}
\end{array} \right] ,~
N_D^{\prime} = \left[ \begin{array}{c}
N_1^{\prime} \\
i \sigma^2 N_2^{\prime *}
\end{array} \right].
\end{eqnarray}
Then, for the small mixing limit, analogy to Ref.~\cite{Baker:2017zwx}, we have new decay channel of 
\begin{equation}
\Gamma(\phi \to\chi\chi)=\frac{(y_\chi\sin\theta)^2}{8\pi m_\phi(T)^2}(m_\phi^2(T)-4m_\chi^2)^{3/2}\;,
\end{equation}
and the dominant annihilation is the $\phi$ mediate s-channel process,
with cross section being given by,
\begin{align}
  \sigma(\phi\phi \to \chi \bar{\chi}) &=
   ( y_\chi\sin\theta)^2 ( \lambda_{\phi} v_\phi(T))^2
    \frac{(s - 4 m_\chi^2)^{3/2}}{8 \pi s (m_\phi^2(T) - s)^2 \sqrt{s - 4 m_\phi^2(T)}} \,,
\\[0.2cm]
  \sigma(H^\dagger H \to \chi \bar\chi) &= 
    4( y_\chi\sin\theta)^2( \lambda_{h\phi } v_\phi(T))^2
    \frac{(s - 4 m_\chi^2)^{3/2}}{8\pi s (m_\phi^2(T) - s)^2 \sqrt{s - 4 m_h^2(T)}} \,.
\end{align}
 
After the vacuum transit from the $\phi$-vacuum to the true Electroweak $h$-vacuum, these channels are shut down, and the decay channel is replaced by $\phi\to \chi N$, and the annihilation channel are mostly from the $t/u$ channel $HH (\phi\phi)\to \chi\bar{\chi}$. We comment that due to the mixing angle $\theta$ is very small, for the FIMP production of $\chi$ with a $y_\chi\sim 10^{-12}$, these contributions are negligible.

\newpage
\bibliography{RHNDM_Relic}
\end{document}